\documentclass[a4paper,aps,prd,reprint,superscriptaddress,nofootinbib,notitlepage,showpacs,showkeys,twocolumn,floatfix]{revtex4-1}

\usepackage[latin2]{inputenc}
\usepackage{amssymb,amsmath,amsfonts}
\usepackage{mathrsfs}
\usepackage{dsfont}
\usepackage{graphicx,bm}
\usepackage{multirow}
\usepackage{color}
\usepackage{t1enc}
\usepackage{enumerate}

\newcommand{\tr}{\mathop{\mathrm{Tr}}}
\newcommand{\x}{\mathbf{x}}

\newcommand{\nom}{{\nonumber}}

\newcommand{\eref}[1]{Eq.~\eqref{#1}}
\newcommand{\secref}[1]{Sec.~\ref{#1}}
\newcommand{\apref}[1]{Appendix~\ref{#1}}
\newcommand{\tabref}[1]{Table~\ref{#1}}

\providecommand{\e}[1]{\ensuremath{{\scriptscriptstyle E\negthinspace #1}}}

\begin{document}
\allowdisplaybreaks

\title{Existence of the critical endpoint in the vector meson extended linear sigma model}

\author{P. Kov\'acs} 
\email{kovacs.peter@wigner.mta.hu}
\affiliation{Institute for Particle and Nuclear Physics, Wigner
  Research Centre for Physics, Hungarian Academy of Sciences, H-1525
  Budapest, Hungary}

\author{Zs. Sz{\'e}p} 
\email{szepzs@achilles.elte.hu}
\affiliation{MTA-ELTE Statistical and Biological Physics Research
  Group, H-1117 Budapest, Hungary}

\author{Gy. Wolf} 
\email{wolf.gyorgy@wigner.mta.hu}
\affiliation{Institute for Particle and Nuclear Physics, Wigner
  Research Center for Physics, Hungarian Academy of Sciences, H-1525
  Budapest, Hungary}

\begin{abstract} 
  The chiral phase transition of the strongly interacting matter is
  investigated at nonzero temperature and baryon chemical potential
  ($\mu_B$) within an extended $(2+1)$ flavor Polyakov constituent
  quark-meson model that incorporates the effect of the vector and
  axial vector mesons. The effect of the fermionic vacuum and thermal
  fluctuations computed from the grand potential of the model is taken
  into account in the curvature masses of the scalar and pseudoscalar
  mesons. The parameters of the model are determined by comparing
  masses and tree-level decay widths with experimental values in a
  $\chi^2$-minimization procedure that selects between various
  possible assignments of scalar nonet states to physical
  particles. We examine the restoration of the chiral symmetry by
  monitoring the temperature evolution of condensates and the chiral
  partners' masses and of the mixing angles for the pseudoscalar
  $\eta-\eta'$ and the corresponding scalar complex. We calculate the
  pressure and various thermodynamical observables derived from it and
  compare them to the continuum extrapolated lattice results of the
  Wuppertal-Budapest collaboration. We study the $T-\mu_B$ phase
  diagram of the model and find that a critical endpoint exists for
  parameters of the model, which give acceptable values of $\chi^2.$
\end{abstract}


\maketitle

\section{Introduction}

We investigate properties of the strongly interacting matter at high
temperature and/or density. Currently, the strong matter can be
accessed experimentally at low density (RHIC/Brookhaven and LHC/CERN)
and at normal nuclear density (ordinary nuclear physics). Its
properties at high densities, where the critical endpoint (CEP)
probably sits, are not known, neither experimentally nor
theoretically. The theory of the strongly interacting matter (QCD) can
be solved perturbatively only at very high energies, not relevant for
the problems here. Lattice computations based on importance sampling
face serious difficulties at finite, especially large
density. Therefore, we are left with effective models, in which
certain aspects of the strongly interacting matter can be studied. The
underlying principle in the construction of such models is that they
share the same global symmetries as the QCD. There are different ways
in which the chiral symmetry can be realized. At large temperatures
and densities, one expects the chiral symmetry of QCD to be
restored. Then, chiral partners have to become degenerate in mass,
e.g., the sigma meson and the pions. To investigate the mechanism of
chiral symmetry restoration, effective theories with linearly realized
chiral symmetry are most appropriate.

In \cite{elsm_2013} an extended linear sigma model (EL$\sigma$M) with
$U(3)_L\times U(3)_R$ global symmetry was developed, which
incorporates the vector and axial vector mesons. The parametrization
of the EL$\sigma$M performed at vanishing temperature shows that the
scalar states are preferred as $\bar q q$ states only if their masses
are above 1~GeV with an opposite ordering $m_{a_0} < m_{K^\star_0}$
compared to the corresponding experimental values.  QCD sum rule
analyses based on Borel transformed two-point correlation functions of
$\bar q q$ currents also predict the masses of $\sigma \equiv
f_0^{\text{L(ow)}}$ (the scalar particle with nonstrange quark
content)\footnote{From now on we always use $f_0^L$ instead of
  $\sigma,$ as in the Particle Data Group (PDG) \cite{PDG}.} 
and $a_0$ to be around 1.2~GeV and larger masses for $K^\star_0$ and
the other $f_0$ ---the $f_0^{\text{H(igh)}}$ ---of the nonet, due to
the strange quark content of the latter (for details see
\cite{Chen:2007xr} and references therein). Only when the above QCD
sum rule analysis is done with tetraquark currents are the masses of
scalar mesons obtained in the region $0.6-1.0$~GeV with the ordering
$m_{f_0^L} < m_{K^\star_0} < m_{f_0^H,a_0}$
\cite{Chen:2007xr,Chen:2009gs,Kojo:2008hk}.

Since the mass of the $f_0^L$, the excitation of the vacuum with
quantum numbers $J^{PC} = 0^{++},$ is intimately related to the
nonstrange condensate, one could expect in the context of the
EL$\sigma$M that a parametrization leading to a large $f_0^L$ mass
will result in a high pseudocritical temperature.  This is because in
the case of a smooth crossover phase transition the larger the $f_0^L$
mass compared to the mass of its chiral partner, that is the pion, the
larger is the temperature at which $m_{f_0^L}$ approaches $m_\pi$ in
the process of the chiral symmetry restoration during which the value
of the nonstrange condensate diminishes.  Another problem with a large
$f_0^L$ mass when the thermodynamics of the EL$\sigma$M is studied
comes from the fact that usually the first order phase transition,
which occurs at $T=0$ as the baryon chemical potential is increased,
weakens with increasing values of the $f_0^L$ mass and eventually
becomes a crossover at a high enough value of $m_{f_0^L}$
\cite{Schaefer:2008hk,Chatterjee:2011jd}.  All this suggests that even
if a zero temperature analysis, which excludes the $f_0^L$ and $f_0^H$
scalar mesons from the parametrization process favors the heavy
scalars as $\bar q q$ states, the combined zero and finite temperature
analysis can give a different result in a given approximate solution
of the model. To completely clarify this issue, it seems necessary to
include in the model all the physical scalar states below 2~GeV, which
is a task we plan to do in a later work along the line of
\cite{Giacosa:2006tf}.

Beside the restoration of the chiral symmetry, the liberation of quarks also
occurs in QCD at high temperature and/or density.  The order parameter of
this deconfinement phase transition in the pure gauge theory is the Polyakov
loop.  It is therefore reasonable to include it in our model in the hope
(supported by existing results in the literature) that in this way a better
phenomenological description of the strongly interacting matter can be
achieved.

We shall study the thermodynamics of the ($2+1$) flavor Polyakov quark
meson model in which, beyond the vector and axial vector mesons
included alongside the scalar and pseudoscalar ones, we take into
account, as fermionic degrees of freedom, the constituent quarks
propagating on a constant gluon background in the temporal direction,
which naturally leads in a mean-field treatment to the appearance of
the Polyakov loop. The influence of the fermionic vacuum fluctuations
on the thermodynamics of the Polyakov loop extended quark meson (PQM)
model proved to be very important. In the case of two flavors
($N_f=2$) it was shown in \cite{Skokov:2010sf} that their inclusion
can change the order of the phase transition at vanishing baryon
chemical potential $\mu_B$ and that renormalization is required to
guarantee the second order nature of the temperature driven phase
transition in the chiral limit. In the PQM model the effect of the
fermionic vacuum fluctuations on the $T-\mu_B$ phase diagram was
investigated, e.g., in \cite{Gupta:2011ez} for $N_f=2$ and in
\cite{Chatterjee:2011jd, Schaefer:2011ex} for $N_f=2+1.$ We shall
incorporate the vacuum fluctuations of the fermions in the grand
potential and study the effect of the inclusion of the (axial) vector
mesons by comparing thermodynamic quantities and the $T-\mu_B$ phase
diagram with those determined in the literature in the context of the
PQM model.

For $N_f=2$ and without the inclusion of fermions, the restoration of
chiral symmetry at high temperature was studied within the EL$\sigma
M$ in Ref.~\cite{Eser:2015pka}, using the functional renormalization
group approach, and in the gauged version of the model in
\cite{Struber:2007bm}, using the Cornwall-Jackiw-Tomboulis formalism
\cite{Cornwall:1974vz}.  An application of the $(2+1)$-flavor
EL$\sigma M$ to an in-medium study was reported in
\cite{Tawfik:2014gga}. In contrast to this latter reference, in which
it is also rather obscure how thermal corrections are included in the
mass of the \hbox{(axial-)}vectors, in the present work we properly
take into account the wave function renormalization factors (neglected
in \cite{Tawfik:2014gga}), which are related to the redefinition of
the \hbox{(axial-)}vector fields and use a complete set of parameters
obtained from a consistent parametrization of the model.

The paper is organized as follows. In Sec.~\ref{Sec:model} we
introduce the model, giving the Lagrangian and the Polyakov loop
potentials considered in this study. In Sec.~\ref{sec:grand_pot} we
introduce the grand potential, the approximation used for its
computation, summarize the determination of the curvature masses and
of the renormalization of the fermion vacuum fluctuations and present
the field equations to be solved numerically. The determination of the
model parameters, which is based on a $\chi^2$-minimization procedure,
is described in detail in Sec.~\ref{Sec:parametrization}. In
Sec.~\ref{Sec:result} we present our results concerning the medium
mass variation of the model constituents, the thermodynamics quantities
derived from the pressure, and the $T-\mu_B$ phase diagram. We conclude
in Sec.~\ref{Sec:conclusion}.\\

\section{Formulation of the Model}
\label{Sec:model}

In this section we give the Lagrangian of the model, introduce the
Polyakov loop, and present the different forms of the Polyakov loop
potential we shall use later. We work with a modified version of the
chiral Lagrangian rather than the one employed in \cite{elsm_2013} at zero
temperature (more details on the construction of chiral Lagrangians
can be found in \cite{Geffen_1969, Kaymakcalan_1984, Ko_1994}). We
consider now a different $U_{A}(1)$ anomaly term (term with $c_1$),
because this term contains the fields with lower powers than the one
used in \cite{elsm_2013}, while it does not affect the zero
temperature properties much (see \cite{U1A_analysis}). Moreover, we
introduce additional kinetic and Yukawa coupling terms for the
constituent fermions $\Psi = (q_u, q_d, q_s)^{T}$. Another important
modification is the presence of the gluon field in the covariant
derivative of the quark field. In the mean-field approximation, this
will give rise in the grand potential of Sec.~\ref{sec:grand_pot} to
the appearance of the Polyakov loop, which mimics some properties of
the quark confinement. Moreover, since $2\to 2$ (axial) vector
scattering processes will not be considered here, the purely four
field (axial) vector self-interaction terms are left out (see
\cite{elsm_2013} for the complete Lagrangian).

\subsection{Lagrangian of the PQM with (axial) vector mesons}

According to the considerations above, the Lagrangian we shall use has
the following form:
\begin{widetext}
\begin{align}
  \mathcal{L} & = \tr[(D_{\mu}M)^{\dagger}(D_{\mu}M)] -
  m_{0}^{2}\tr(M^{\dagger}M) - \lambda_{1}[\tr(M^{\dagger} M)]^{2} - \lambda_{2}\tr(M^{\dagger}M)^{2} 
  + c_{1}(\det M+\det M^{\dagger}) + \tr[H(M+M^{\dagger})] \nom \\
  & -\frac{1}{4}\tr(L_{\mu\nu}^{2}+R_{\mu\nu}^{2}) 
  + \tr\left[ \left(\frac{m_{1}^{2}}{2}+\Delta\right) (L_{\mu}^{2}+R_{\mu}^{2})\right] 
  + i\frac{g_{2}}{2}(\tr\{L_{\mu\nu}[L^{\mu},L^{\nu}]\} + \tr\{R_{\mu\nu}[R^{\mu},R^{\nu}]\})\nom \\
  & +\frac{h_{1}}{2}\tr(M^{\dagger}M)\tr(L_{\mu} ^{2}+R_{\mu}^{2})
  + h_{2}\tr(|L_{\mu}M|^{2}+|M R_{\mu} |^{2}) + 2h_{3}\tr(L_{\mu}M R^{\mu}M^{\dagger}) 
    + \bar{\Psi}\left[i \gamma_{\mu}D^{\mu}-\mathcal{M}\right]\Psi\, . 
\label{Eq:Lagr}
\end{align}
The covariant derivatives appearing in \eqref{Eq:Lagr} are written in
terms of the electromagnetic field $A_{e}^{\mu}$, the left- and
right-handed vector fields $L^{\mu}, R^{\mu}$ and the gluon fields
$G^{\mu}_i$ as
\begin{eqnarray}
  D^{\mu}M = \partial^{\mu}M-i g_{1}(L^{\mu}M-M R^{\mu})-i e
  A_{e}^{\mu}[T_{3},M], \quad D^{\mu}\Psi = \partial^{\mu}\Psi - i
  G^{\mu}\Psi,\label{eq:psi_partial}
\end{eqnarray} 
where $G^{\mu} = g_s G^{\mu}_i T_i,$ with $T_{i} = \lambda_{i}/2$
($i=1,\ldots,8$) denoting the $SU(3)$ group generators given in terms
of the Gell-Mann matrices $\lambda_{i}.$ The field strength tensors
\begin{eqnarray}
L^{\mu\nu} = \partial^{\mu}L^{\nu} - ieA_{e}^{\mu}[T_{3},L^{\nu}] -
\left\{\partial^{\nu}L^{\mu} - ieA_{e}^{\nu}[T_{3},L^{\mu}]\right\},\quad
R^{\mu\nu} = \partial^{\mu}R^{\nu} - ieA_{e}^{\mu}[T_{3},R^{\nu}] -
\left\{ \partial^{\nu}R^{\mu}-ieA_{e}^{\nu}[T_{3},R^{\mu}]\right\},
\end{eqnarray} 
are constructed from the left- and right-handed vector fields
$L^{\mu}$ and $R^{\mu}$ which contain the nonets of vector
($V_{a}^{\mu}$) and axial vector ($A_{a}^{\mu}$) meson fields as
follows:
\begin{align}
L^{\mu}  &  \equiv V^{\mu} + A^{\mu} \equiv
\sum_{a=0}^{8}(V_{a}^{\mu}+A_{a}^{\mu})T_{a} = \frac{1}{\sqrt{2}
}\left(
\begin{array}
[c]{ccc}
\frac{\omega_{N}+\rho^{0}}{\sqrt{2}}+\frac{f_{1N}+a_{1}^{0}}{\sqrt{2}} &
\rho^{+}+a_{1}^{+} & K^{\star+}+K_{1}^{+}\\
\rho^{-}+a_{1}^{-} & \frac{\omega_{N}-\rho^{0}}{\sqrt{2}}+\frac{f_{1N}
-a_{1}^{0}}{\sqrt{2}} & K^{\star0}+K_{1}^{0}\\
K^{\star-}+K_{1}^{-} & {\bar{K}}^{\star0}+{\bar{K}}_{1}^{0} & \omega_{S}+f_{1S}
\end{array}
\right)  ^{\mu},\label{eq:matrix_field_L}\\
R^{\mu}  & \equiv V^{\mu} - A^{\mu} \equiv
\sum_{a=0}^{8}(V_{a}^{\mu}-A_{a}^{\mu})T_{a} = \frac{1}{\sqrt{2}}
\left(
\begin{array}
[c]{ccc}
\frac{\omega_{N}+\rho^{0}}{\sqrt{2}}-\frac{f_{1N}+a_{1}^{0}}{\sqrt{2}} &
\rho^{+}-a_{1}^{+} & K^{\star+}-K_{1}^{+}\\
\rho^{-}-a_{1}^{-} & \frac{\omega_{N}-\rho^{0}}{\sqrt{2}}-\frac{f_{1N}
-a_{1}^{0}}{\sqrt{2}} & K^{\star0}-K_{1}^{0}\\
K^{\star-}-K_{1}^{-} & {\bar{K}}^{\star0}-{\bar{K}}_{1}^{0} & \omega_{S}-f_{1S}
\end{array}
\right)  ^{\mu}, \label{eq:matrix_field_R}
\end{align}
where the assignment to physical fields is made explicit with the
exception of the mixing sector. The index $a=0,\ldots,8$ runs over the
generators of the $U(3)$ group which includes also $T_0=\lambda_0/2$
with $\lambda_0 = \sqrt{\frac{2}{3}} \mathds{1}_{3\times 3}.$ The
matrix $M$ in the Lagrangian collects the nonets of scalar ($S_{a}$)
and pseudoscalar ($P_{a}$) meson fields, 
\begin{equation}
  M\equiv M_{S} + M_{PS}
  \equiv \sum_{a=0}^{8}(S_{a}+iP_{a})T_{a}=\frac{1}{\sqrt{2}}\left(
    \begin{array}
      [c]{ccc}%
      \frac{(\sigma_{N}+a_{0}^{0})+i(\eta_{N}+\pi^{0})}{\sqrt{2}} & a_{0}^{+}
      +i\pi^{+} & K_{0}^{\star+}+iK^{+}\\
      a_{0}^{-}+i\pi^{-} & \frac{(\sigma_{N}-a_{0}^{0})+i(\eta_{N}-\pi^{0})}
      {\sqrt{2}} & K_{0}^{\star0}+iK^{0}\\
      K_{0}^{\star-}+iK^{-} & {\bar{K}_{0}^{\star0}}+i{\bar{K}^{0}} & \sigma_{S}+i\eta_{S}
\end{array}
\right), \label{eq:matrix_field_Phi}
\end{equation}
\end{widetext}
while the external fields related to the scalar and vector fields are
introduced with the following parametrization:
\begin{align}
H  & =H_{0}T_{0}+H_{8}T_{8}=
\frac{1}{2}\textnormal{diag}(h_{0N},h_{0N},\sqrt{2}h_{0S}),
\label{eq:expl_sym_br_epsilon}\\
\Delta & = \Delta_{0}T_{0}+\Delta_{8}T_{8}=
\textnormal{diag}(\delta_{N}, \delta_{N}, \delta_{S}).
\label{eq:expl_sym_br_delta}
\end{align}

The first line in the Lagrangian \eqref{Eq:Lagr} contains the kinetic
and self-interaction terms of the (pseudo)scalars together with a
$U_{A}(1)$ anomaly term and an explicit symmetry breaking term. The
second line consists of the kinetic terms for the (axial)vectors,
altogether with explicit symmetry breaking terms for the
(axial)vectors and the (axial)vector--electromagnetic interaction
terms. In the third line one finds the (pseudo)scalar--(axial)vector
interaction terms, the kinetic terms of the constituent quarks and
their Yukawa-type interaction with the (pseudo)scalar mesons. The
quark mass matrix appearing there is defined as
\begin{equation}
\mathcal{M} = g_F\left(\mathds{1}_{4\times 4} M_{S} + i\gamma_5 M_{PS}\right),
\label{Eq:q_mass_matrix}
\end{equation} 
and has the structure of a block matrix in flavor, Dirac, and color space.

For convenience, in the matrices above and throughout the article, we
use the $N-S$ (nonstrange--strange) basis instead of the $0-8$ basis,
which for a generic field $\xi_a\in(S_a,P_a,V_a^{\mu},A_a^{\mu}, H_a,
\Delta_a)$ is defined as
\begin{equation}
\xi_N = \frac{1}{\sqrt{3}} \left( \sqrt{2}\; \xi_0+ \xi_8\right), \quad 
\xi_S = \frac{1}{\sqrt{3}} \left( \xi_0-\sqrt{2}\;\xi_8 \right).   
\label{eq:nsbase}
\end{equation}
Since in the present work we neglect the isospin breaking, we have to
deal with only two nonzero condensates (field expectation values), the
$\phi_N \equiv \langle \sigma_N \rangle$ nonstrange and the
$\phi_S \equiv \langle \sigma_S \rangle$ strange scalar condensates.
In the broken symmetry phase, the model Lagrangian is obtained with the
usual procedure in which the nonstrange and strange scalar fields are
shifted by their expectation values, $\sigma_{N/S}\to \sigma_{N/S} + \phi_{N/S}$,
which will generate the tree-level masses and decay widths.

\subsection{The Polyakov loop potential}

The introduction of the Polyakov loop operator and its application in
the present context can be found, for instance, in
\cite{Holland:2000uj, Fukushima:2003fw, Ratti:2005jh}. For the sake of
completeness, however, let the key steps be presented here as well.

To go to finite temperature, analytic continuation to imaginary time
should be performed, $t \to -i\tau$. The temporal component of the
gluon gauge field, which is entering in the definition of the Polyakov
loop operator, is transformed accordingly as $G_0(t, \x) \to -iG_4(\tau,
\x)$, while we assume that the spatial components of $G^\mu$ are
vanishing. The Polyakov loop operator itself---which is nothing other
than a path ordered Wilson loop of the gauge field in the temporal
direction---is defined as \cite{Chatterjee:2011jd, Hansen:2006ee}
\begin{equation}
L = \mathcal{P}\exp\left(i\int_{0}^{\beta}d\tau G_4(\tau,\x) \right).
\end{equation}
$L$ and $L^\dagger$ are matrices in the fundamental representation of
the color gauge group $SU(N_c)$ with $N_c=3$. Introducing the color
traced Polyakov loops as
\begin{equation}
\Phi(\x) = \frac{1}{N_c}\tr\nolimits_c L(\x),
\quad \bar\Phi(\x) = \frac{1}{N_c} \tr \nolimits_c L^{\dagger} (\x)\,,
\label{Eq:Phi_barPhi}
\end{equation}
the Polyakov loop variables are defined as the thermal expectation
values $\langle \Phi\rangle$ and $\langle \bar \Phi\rangle$. In the
pure gauge case they are related to the free energy of infinitely
heavy static quark and antiquarks.

As a next step, the so-called Polyakov gauge is chosen, in which
$G_4(\tau,\x)=G_4(\x)$ is time independent and diagonal in color
space; that is, it belongs to the Cartan subalgebra. Furthermore, we
approximate $G_4(\x)$ to be homogeneous, thus it can be written as
\begin{equation}
G_4 = \phi_3\lambda_3+\phi_8\lambda_8 \label{eq:G4_form}\,,
\end{equation}
with $\phi_3$ and $\phi_8$ being real. Consequently,
with these simplifications the Polyakov loop operator can be cast into
the following form: 
\begin{equation}
L=\mathrm{diag}(z_1,z_2,z_1^{-1}z_2^{-1}),\label{eq:L_form}
\end{equation}
with
$z_1=e^{i\beta(\phi_3+\phi_8/\sqrt{3})}, z_2=e^{i\beta(-\phi_3+\phi_8/\sqrt{3})}$. 
When the constant diagonal $G_4,$ given in \eqref{eq:G4_form}, is
substituted into the kinetic term of the constituent quarks
\eqref{eq:psi_partial}, the second term of the covariant derivative
can be considered as a color dependent imaginary chemical
potential. This observation is used for the calculation of the
grand canonical potential in \secref{sec:grand_pot}.

The Polyakov loop potential describes the temperature driven
deconfinement phase transition occurring in the pure gauge theory;
therefore, the potential is constructed using terms which are
invariant under the $Z(3)$ symmetry, and some coefficient of these
terms depend on the temperature in order to assure a nonzero
expectation value of $\Phi$ at large temperature
\cite{Pisarski_2000,Scavenius:2002ru}. The potential is constructed in
such a way as to reproduce some thermodynamical quantities of the pure
gauge theory computed on the lattice. For the functional form there
are still various possibilities. The simplest polynomial potential
introduced in \cite{Pisarski_2000} leads in Polyakov
Nambu-Jona-Lasinio (PNJL) or PQM models to some unwanted properties,
such as negative susceptibilities \cite{Sasaki_2006}.  Therefore, we
shall use a potential with a logarithmic form which is coming from the
$SU(3)$ Haar measure of the group integration \cite{Roessner_2006} and
is free from the negative susceptibility problem. Moreover, as
observed in \cite{Haas:2013qwp}, the trace anomaly calculated with the
logarithmic parametrization of the Polyakov loop potential shows a
better agreement with the corresponding quantity in the pure $SU(3)$
gauge theory computed recently on the lattice in
\cite{Borsanyi:2012ve}, compared to the case when a polynomial
Polyakov loop potential is used.

Although in thermodynamical applications the potential is a function of
the expectation values $\langle \Phi\rangle$ and $\langle \bar
\Phi\rangle,$ we use for simplicity $\Phi$ and $\bar\Phi$ for its
arguments. Then the logarithmic Polyakov loop potential can be written
as
\begin{align}
&\beta^4U_{\text{log}}(\Phi,\bar{\Phi}) =
-\frac{1}{2}a(T)\Phi\bar{\Phi} \nonumber \\
&\qquad + b(T)\ln\left(1 - 6\Phi\bar{\Phi} +
  4(\Phi^3 + \bar{\Phi}^3) - 3(\Phi\bar{\Phi})^2\right), \label{eq:Ulog}
\end{align}
with coefficients
\begin{equation}
a(T) = a_0 + a_1\left(\frac{T_0}{T}\right) +
a_2\left(\frac{T_0}{T}\right)^2, \quad b(T) = b_3\left(\frac{T_0}{T}\right)^3,
\end{equation} 
where the values of the constants are $a_0=3.51$, $a_1=-2.47$,
$a_2=15.22$, and $b_3=-1.75$.  

The above parametrization of the Polyakov loop potential does not include
the backreaction of the dynamical quarks on the gauge sector and therefore
the influence of the quarks on the deconfinement phase transition.  This
effect was discussed in \cite{Schaefer:2007pw} and in \cite{Haas:2013qwp},
where the dependence of $T_0$ on the number of quark flavors and the baryon
chemical potential was estimated.  This led to $T_0=187$~MeV for
$m_s=150$~MeV and $T_0=182$~MeV for $m_s=95$~MeV.  In the present study we
shall use this latter value of $T_0.$ A refinement of this estimation was
achieved in \cite{Haas:2013qwp}, where a quark-improved Polyakov loop
potential was constructed by comparing the $SU(3)$ Yang-Mills (YM) effective
potential with the gluonic effective potential computed with the functional
renormalization group method by including the quark polarization in the
gluon propagator.  It was observed that the two potentials have the same
shape and that they can be mapped into each other by relating the
temperatures of the two systems, $T_{\text{YM}}$ and $T_{\text{glue}},$
respectively.  The use of the improved Polyakov loop potential
$U_{\text{glue}}$ was proposed in \cite{Haas:2013qwp}, which, denoting by
$U_\text{YM}$ the potentials in \eqref{eq:Ulog}, was constructed based on
the relation \begin{equation}
\frac{1}{T^4_\text{glue}}\big[U_\text{glue}(\Phi,\bar\Phi)\big]
\big|_{t_\text{glue}} = \frac{1}{T^4_\text{YM}}
\big[U_\text{YM}(\Phi,\bar\Phi) \big]\big|_{t_\text{YM}(t_\text{glue})},
\label{Eq:U_glue} \end{equation} where the mapping between the reduced
temperatures $t_\text{YM}=T_\text{YM}/T^\text{YM}_\text{c}-1$ and
$t_\text{glue}=T_\text{glue}/T^\text{glue}_\text{c}-1$ was determined to be
$t_\text{YM}(t_\text{glue})\approx0.57\, t_\text{glue},$ with the critical
temperatures $T^\text{YM}_\text{c}=270$~MeV and
$T^\text{glue}_\text{c}\in[180,270]$~MeV.  In practice this amounts to using
in the right-hand side of \eqref{eq:Ulog}, where $T_0$ means
$T^\text{YM}_\text{c},$ the replacement $T\to T_\text{c}^\text{YM} (1 + 0.57
(T/T_\text{c}^\text{glue} - 1))$ (on the left side of the arrow $T\equiv
T_{\text{YM}},$ while on the right side $T\equiv T_{\text{glue}}$).  In
Sec.~\ref{Sec:result} we shall choose several values of
$T_\text{c}^\text{glue}$ in the range given above and study the sensitivity
of the results to this parameter.

Before closing this section we mention that a gluonic potential with
possible phenomenological applicability is also calculated in
\cite{Reinosa:2015gxn} in terms of the Polyakov loop variables
$\langle \Phi\rangle$ and $\langle \bar \Phi\rangle,$ using background
field methods in the massive extension of the Landau-deWitt gauge.

\section{The Grand Potential}
\label{sec:grand_pot}

To study the thermodynamics of a symmetric quark matter
($\mu_u=\mu_d=\mu_s\equiv\mu_q=\mu_B/3$), we shall use the grand
potential $\Omega(T,\mu_q)$ obtained from the partition function of a
three-dimensional spatially uniform system of volume $V$ in thermal
equilibrium at temperature $T=1/\beta.$ Following Ref.~
\cite{Kapusta:2006pm} the partition function can be given the
following representation in terms of path integrals:
\begin{align}
{\mathcal{Z}} &= e^{-\beta V \Omega(T,\mu_q)}=\textnormal {Tr}\exp
\bigg[ -\beta \bigg( \hat {\mathcal{H}} - \sum_{f=u,d,s}\mu_f \hat
{\mathcal{Q}}_f \bigg)\bigg]\nonumber \\
&=\int\limits_\textnormal{PBC} \prod_a {\mathcal{D}} \xi_a
\int\limits_\textnormal{APBC} \prod_f {\mathcal{D} q_f} {\mathcal{D} q^\dagger_f} \nonumber \\
&\ \ \times \exp\bigg[-\int_0^\beta d\tau \int_V d^3 x \bigg(
{\mathcal{L}} + \mu_q\sum_{f} q^\dagger_f q_f\bigg)\bigg],
\end{align}
where (A)PBC stands for (anti)periodic boundary condition,
$\hat{\mathcal{Q}}_f$ is the conserved charge operator, and $\xi$
denotes here all the mesonic fields. Since the Polyakov loop 
is treated at mean-field level, there is no integration over the
gluons [$G_4$ in \eqref{eq:G4_form} is a background field]
and in this case the Polyakov loop potential \eqref{eq:Ulog} is simply
added to the grand potential.

The simplest approximation for the evaluation of the grand potential
frequently used in the literature takes into account the
(pseudo)scalar mesons at mean-field level only. In the present case
the vacuum and thermal fluctuations for the fermions are taken into
account, while the mesonic vacuum fluctuations are neglected and the
effects of the lightest mesonic thermal fluctuations ($\pi$, $K$,
$f_0^L$) are included only in the pressure and the thermodynamical
quantities derived from it. Therefore, the meson potential is
classical (tree-level) and the fermion determinant obtained after
performing the functional integration over the quark fields is
evaluated for vanishing mesonic fluctuating fields. Since we would
like to assess how the parametrization using vector and axial vector
mesons influences the thermodynamics in this approximation, we shall
also neglect the fluctuations of the vector and axial vector
mesons. In this approximation, which we shall call {\it hybrid} (H)
approximation, the grand potential reads
\begin{equation}
\Omega_\textnormal{H}(T,\mu_q) = U(\left<M\right>) + 
U(\langle\Phi\rangle,\langle\bar\Phi\rangle)
+ \Omega_{\bar q q}^{(0)}(T,\mu_q),
\label{Eq:grand_pot_H}
\end{equation}

where $U(\left<M\right>)$ is the tree-level meson potential,
$U(\langle\Phi\rangle,\langle\bar\Phi\rangle)$ is the Polyakov loop
potential and $\Omega_{\bar q q}^{(0)}$ is the contribution of the
fermions for nonvanishing scalar backgrounds $\phi_N$ and $\phi_S$ and
vanishing mesonic fluctuating fields, the case in which the quark mass
matrix $\mathcal{M}$ given in \eqref{Eq:q_mass_matrix} is diagonal in
flavor space. Note that $\mathcal{M}$ has a nontrivial dependence on
the scalar and pseudoscalar fluctuating fields, when they are
nonvanishing.

The tree-level mesonic potential
\begin{align}
  \!\!\! U(\left<M\right>)&=\frac{m_0^2}{2}\big(\phi_N^2+\phi_S^2\big)
  -\frac{c_1}{2\sqrt{2}}\phi_N^2\phi_S - h_S\phi_S - h_N\phi_N  \nonumber\\
  &+\frac{\lambda_1}{4}\big(\phi_N^2+\phi_S^2\big)^2+\frac{\lambda_2}{8}\big(\phi_N^4+2\phi_S^4\big)\nonumber \\
  &+\frac{\delta m_0^2}{2}\big(\phi_N^2+\phi_S^2\big) +
  \frac{\delta\lambda_2}{8}\big(\phi_N^4+2\phi_S^4\big),
\label{Eq:meson_pot}
\end{align}
is obtained from the first line of \eqref{Eq:Lagr} with the
replacement $M,M^\dagger\to \left<M\right> \equiv T_N\phi_N +
T_S\phi_S,$ where $T_{N/S}=\lambda_{N/S}/2$ with $\lambda_N =
\textnormal{diag}(1,1,0)$ and $\lambda_S = \textnormal{diag}(0, 0,
\sqrt{2}).$ In the last line of \eqref{Eq:meson_pot} we explicitly
added the counterterms which are needed to renormalize the fermionic
vacuum fluctuations (see Sec.~\ref{Subsec:ren}).

The contribution of the fermions to the grand potential in the
approximation described above is obtained as
\begin{align}
&\!\!\!{\mathcal{Z}_{\bar q q}^{(0)}}  = e^{-\beta V \Omega_{\bar q q}^{(0)}} =
  \int\limits_\textnormal{APBC} \prod_f {\mathcal{D} q_f} {\mathcal{D}
    q^\dagger_f}
  \exp \bigg\{ \int_0^\beta d\tau \int_V d^3 x \nonumber\\
& \times q_f^\dagger \bigg[\bigg(i\gamma_0 \vec\gamma\cdot\vec\nabla
  -\frac{\partial}{\partial \tau}+\tilde \mu_q\bigg)\delta_{f g} -
  \left.\gamma_0 {\mathcal{M}}_{fg}\right|_{\xi_a=0} \bigg]
  q_g\bigg\},
\label{Eq:Zqq0}
\end{align}
where summation over repeated indices $f,g\in\{u,d,s\}$ is
understood, the superscript $(0)$ reminds one that the mesonic fluctuating
fields $\xi_a$ are set to zero in the quark mass matrix $\mathcal{M}$
defined in \eqref{Eq:q_mass_matrix} and we introduced the
color-dependent chemical potential $\tilde \mu_q=\mu_q-i G_4,$
different for each color.

Evaluating the path integral in \eqref{Eq:Zqq0} as in
\cite{Kapusta:2006pm} one obtains
\begin{equation} 
  \Omega_{\bar q q}^{(0)}(T,\mu_q) = \Omega_{\bar q q}^{(0)\textnormal{v}} + \Omega_{\bar q q}^{(0)\textnormal{T}}(T,\mu_q),
\end{equation}
where the vacuum and thermal parts are, respectively,
\begin{align}
  &\Omega_{\bar q q}^{(0)\textnormal{v}} = -2 N_c \sum_{f=u,d,s} \int\frac{d^3 p}{(2\pi)^3} E_f(p), \label{Eq:fermion_vac}\\
  &\Omega_{\bar q q}^{(0)\textnormal{T}}(T,\mu_q)  = -2 T \sum_{j=1}^{N_c} \sum_{f=u,d,s} \int\frac{d^3 p}{(2\pi)^3} \nonumber \\
  &\times \big[\ln\big(1+ e^{-\beta (E_f(p)-\tilde \mu_q^j)}\big)
  +\ln\big(1+ e^{-\beta (E_f(p)+\tilde \mu_q^j)}\big)\big]. \ \label{Eq:fermion_thermal}
\end{align}
Here $\tilde \mu_q^j = \mu_q - i (G_4)_{jj},$
$E_f(p)=\sqrt{p^2+m_f^2}$ with $p=|{\bf p}|$ and, in the 
nonstrange--strange basis, the constituent quark masses are given by
\begin{equation}
  m_{u,d} = \frac{g_F}{2}\phi_N \quad \textrm{and}
  \quad m_s = \frac{g_F}{\sqrt{2}} \phi_S.
\label{Eq:uds_masses}
\end{equation}
Writing
\begin{equation}
\sum_{j=1}^{N_c} \ln\big(1 + e^{-\beta (E_f(p)\mp \tilde \mu_q^j)}\big) =
\textnormal{Tr}_c \ln\big(1 + e^{\mp i\beta G_4} e^{-\beta
E_f^{\pm}(p)}\big),
\label{Eq:color-sum}
\end{equation}
one recognizes the appearance of $L=e^{i\beta G_4}$ and 
$L^\dagger=e^{-i\beta G_4},$ given explicitly in \eqref{eq:L_form}. 
Using the properties $\det L = \det L^\dagger =1$ and $L^\dagger L=1$ 
one expresses \eqref{Eq:color-sum} in terms of $\Phi=\textnormal{Tr}_cL/3$ 
and $\bar\Phi=\textnormal{Tr}_c L^\dagger/3$. We obtain
\begin{equation}
  \Omega_{\bar q q}^{(0)\textnormal{T}}(T,\mu_q) = -2 T \sum_f\int
\frac{d^3 p}{(2\pi)^3}\big[\ln g_f^+(p) + \ln g_f^-(p)\big],\ \ 
\label{Eq:fermi_omega}
\end{equation}
where $\Phi^+ = \bar\Phi$ and $\Phi^-=\Phi$ were introduced for
convenience in order to write in a compact form
\begin{equation}
  g_f^{\pm}(p) = 1 + 3\left( \Phi^\pm + \Phi^\mp e^{-\beta E_f^\pm(p)}
  \right) e^{-\beta E_f^\pm(p)} + e^{-3\beta  E_f^\pm(p)},
\label{Eq:g_Phi_barPhi}
\end{equation} 
with $E_f^{\pm}(p) = E_f(p)\mp\mu_f.$

\subsection{Renormalization of the fermionic vacuum contribution}
\label{Subsec:ren}

Using a three-dimensional cutoff $\Lambda$ in the fermionic vacuum
term \eqref{Eq:fermion_vac}, one obtains with the help of the mass
formulas in \eqref{Eq:uds_masses} 
\begin{align}
  &-6 \sum_{f=u,d,s} \int\frac{d^3 p}{(2\pi)^3} E_f(p) \theta(\Lambda
  -
  p) = -\frac{9\Lambda^4}{4\pi^2} - \frac{3g_F^2}{8\pi^2}\Lambda^2\nonumber \\
  &\qquad\qquad \times
  \big(\phi_N^2+\phi_S^2\big)+\frac{3g_F^4}{64\pi^2}\ln\Big(\frac{2\Lambda}{M_0
    e^{\frac{1}{4}}}\Big) \big(\phi_N^4+2\phi_S^4\big)\nonumber \\
  &\qquad\qquad -\frac{3}{8\pi^2}\sum_{f=u,d,s} m_f^4
  \ln\frac{m_f}{M_0} +
  \mathcal{O}\left(\frac{m_f^6}{\Lambda^2}\right).
\end{align}
The first term on the right-hand side, quartic in $\Lambda,$ is
uninteresting and can be removed from the potential by considering a
subtracted potential such that the value of the potential at
$\phi_N=\phi_S=0$ is subtracted. The quadratic and logarithmic
divergences can be removed with the help of the counterterms in the
tree-level mesonic potential \eqref{Eq:meson_pot} by choosing
\begin{equation}
\delta m_0^2 = \frac{3 g_F^2}{4\pi^2}\Lambda^2\quad \textnormal{and}\quad
\delta\lambda_2 = - \frac{3 g_F^4}{8\pi^2} \ln\frac{2\Lambda}{M_0 e^{\frac{1}{4}}}.
\end{equation}
Therefore, the renormalized fermionic vacuum contribution reads 
\begin{equation}
\Omega_{\bar q q;R}^{(0)\textnormal{v}} = -\frac{3}{8\pi^2}\sum_{f=u,d,s} m_f^4 \ln\frac{m_f}{M_0}.
\label{Eq:fermion_vac_ren}
\end{equation}

It was shown in Refs.~\cite{Chatterjee:2011jd,Tiwari:2013pg} that the
grand potential is independent of the renormalization scale, which
means that $d\Omega_\textnormal{H}/dM_0 \equiv 0.$ The reason behind
this is that after renormalization $\lambda_2$ becomes a quantity that
depends on the renormalization scale $M_0$ and its $\beta$ function is
$\beta_{\lambda_2}=\frac{d \lambda_2}{d \ln M_0}= -\frac{3
  g^4_F}{8\pi^2},$ so that the $M_0$ dependence of $\lambda_2$
compensates the explicit dependence on $M_0$ of the renormalized
vacuum term \eqref{Eq:fermion_vac_ren}. As a consequence, we could
freely choose the renormalization scale $M_0$ and maintain $M_0$
independency as long as we take into account the $M_0$ dependence of
$\lambda_2$ (which means we cannot change $M_0$ and $\lambda_2$
independently). However, during the parametrization we scan through
the parameter space uniformly treating all parameters independently;
thus we do not use the $M_0$ dependence of $\lambda_2$, but instead,
we consider $M_0$ as one of the parameters (see
\secref{Sec:parametrization} for additional details).

\begin{table*}[t]
  \caption{
    The components of the pseudoscalar and scalar tree-level mass
    squared matrices and the corresponding contribution of the fermionic
    vacuum and thermal fluctuations given in the $N-S$ basis. We
    introduced $\Lambda_1=\lambda_1+\lambda_2/2,$
    $\Lambda_2=\lambda_1+\lambda_2,$ $\Lambda_3=\lambda_1+3\lambda_2/2,$
    $A=3 g_F^4/(64\pi^2),$ $C=6 g_F^2,$ and following
    \cite{Chatterjee:2011jd} $X=1+4\ln\frac{g_F\phi_N}{2M_0}$ and
    $Y=1+4\ln\frac{g_F\phi_S}{\sqrt{2}M_0},$ with $M_0$ being the
    renormalization scale. $T_f,$ the thermal part of the tadpole
    integral, is defined in \eqref{Eq:tad_th} and $B_f=-d T_f/(d
    m_f^2).$ \label{Tab:m2_matrix}
  } 
\begin{ruledtabular}
\begin{tabular}{lll}
  Tree-level meson squared masses & Fermionic vacuum correction & Fermionic thermal correction \\ \hline
  $\mathfrak{m}_\pi^2=Z_\pi^2(m_0^2+\Lambda_1\phi_N^2+\lambda_1\phi_S^2-c_1\phi_S/\sqrt{2})$ 
  & $\Delta m_\pi^2=-A Z_\pi^2\phi_N^2 X$ & $\delta  m_\pi^2=C Z_\pi^2 T_u$ \\
  $\mathfrak{m}_K^2=Z_K^2[m_0^2+\Lambda_1\phi^2_N+\Lambda_2\phi^2_S-(c_1+\sqrt{2}\lambda_2\phi_S)\phi_N/2]\quad$ 
  & $\Delta m_K^2=-A Z_K^2\frac{\phi_N^3 X + 2\sqrt{2} \phi_S^3 Y}{\phi_N+\sqrt{2}\phi_S}$ 
  & $\delta m_K^2 = C Z_K^2\frac{\phi_N T_u+\sqrt{2}\phi_S T_s}{\phi_N + \sqrt{2}\phi_S}$ \\
  $\mathfrak{m}_{\eta_N}^2=Z_{\eta_N}^2(m_0^2+\Lambda_1\phi_N^2+\lambda_1\phi_S^2+c_1\phi_S/\sqrt{2})$
  & $\Delta m_{\eta_N}^2=-A Z_{\eta_N}^2\phi_N^2 X$ & $\delta m_{\eta_N}^2 = C Z_{\eta_N}^2 T_u$  \\
  $\mathfrak{m}_{\eta_S}^2=Z_{\eta_S}^2(m_0^2+\lambda_1\phi_N^2+\Lambda_2\phi_S^2)$
  & $\Delta m_{\eta_S}^2=-2 A Z_{\eta_S}^2\phi_S^2 Y$ & $\delta m_{\eta_S}^2 = C Z_{\eta_S}^2 T_s$  \\
  $\mathfrak{m}_{\eta_{NS}}^2=Z_{\eta_N} Z_{\eta_S}c_1\phi_N/\sqrt{2}$
  & $\Delta m_{\eta_{NS}}^2 = 0$ & $\delta m_{\eta_{NS}}^2 = 0$\vspace*{0.1cm}\\ \hline
  $\mathfrak{m}_{a_0}^2=m_0^2+\Lambda_3\phi^2_N+\lambda_1\phi^2_S+c_1\phi_S/\sqrt{2}$ 
  & $\Delta m_{a_0}^2=-A\phi_N^2 (4+3 X)$ & $\delta m_{a_0}^2=C (T_u-\frac{g_F^2\phi_s^2}{2}B_u)$  \\
  $\mathfrak{m}_{K_0^\star}^2=Z_{K_0^\star}^2[m_0^2+\Lambda_1\phi^2_N+\Lambda_2\phi^2_S+(c_1+\sqrt{2}\lambda_2\phi_S)\phi_N/2]$
  & $\Delta m_{K_0^\star}^2=-A Z_{K_0^\star}^2\frac{\phi_N^3 X - 2\sqrt{2} \phi_S^3 Y}{\phi_N-\sqrt{2}\phi_S}$ 
  & $\delta m_{K_0^\star}^2=C Z_{K_0^\star}^2\frac{\phi_N T_u-\sqrt{2}\phi_S T_s}{\phi_N - \sqrt{2}\phi_S }$ \\
  $\mathfrak{m}_{\sigma_N}^2=m_0^2+3\Lambda_1\phi_N^2+\lambda_1\phi_S^2-c_1\phi_S/\sqrt{2}$
  & $\Delta m_{\sigma_N}^2=-A\phi_N^2(4+3 X)$ & $\delta m_{\sigma_N}^2=C (T_u-\frac{g_F^2\phi_N^2}{2}B_u)$ \\
  $\mathfrak{m}_{\sigma_S}^2=m_0^2+\lambda_1\phi^2_N+3\Lambda_2\phi_S^2$ & $\Delta m_{\sigma_S}^2=-2 A \phi_S^2(4+3 Y)$ & 
  $\delta m_{\sigma_S}^2=C (T_s-\frac{g_F^2\phi_S^2}{2}B_s)$  \\
  $\mathfrak{m}_{\sigma_{NS}}^2=2\lambda_1\phi_N\phi_S-c_1\phi_N/\sqrt{2}$ & $\Delta m_{\sigma_{NS}}^2=0$ & $\delta m_{\sigma_{NS}}^2=0$
  \vspace*{0.1cm}\\
\end{tabular}
\end{ruledtabular}
\end{table*}

\subsection{The curvature meson masses}

The squared masses of the scalar and pseudoscalar mesons, used later
to determine the parameters of the model, are calculated from the
elements of the corresponding curvature matrix, that is, the second
derivative of the grand potential with respect to the mesonic fields,
generally denoted by $\varphi_{i,a}$ in some appropriate basis indexed
by $a,$ with $i=S$ for scalar and $i=P$ for pseudoscalar mesons. These
curvature matrices are symmetric and nondiagonal in the $0-8$ or
nonstrange--strange basis and can be decomposed as
\begin{equation}
  m^2_{i,{ab}} = \frac{\partial^2 \Omega(T,\mu_q )}{\partial
    \varphi_{i,a} \partial \varphi_{i,b}}
  \bigg|_\textnormal{min}=\mathfrak{m}^2_{i,ab}+\Delta m^2_{i,ab}+\delta
  m^2_{i,ab}, 
\label{Eq:M2i_ab}
\end{equation}
where the three terms on the right-hand side are as follows:
$\mathfrak{m}^2_{i,ab}$ is the tree-level mass
matrix,\footnote{Compared to the case of the conventional L$\sigma$M,
  some elements of this matrix contain the wave-function
  renormalization constants $Z_{\pi}=Z_{\eta_N},$ $Z_K,$ $Z_{\eta_S},$
  $Z_{K_0^{\star}}$, which are needed in order to maintain the
  canonical normalization of the fields in the presence of axial and
  vector mesons (see \cite{elsm_2013} for details).} and $\Delta
m^2_{i,ab}$ and $\delta m^2_{i,ab}$ are the contributions of the
fermionic vacuum and thermal fluctuations, respectively. We note that
the mesonic fields are set to their expectation value only after the
differentiation is performed.

In the case of three flavors, $\delta m^2_{i,ab}$ was first calculated
without the inclusion of the Polyakov loop in \cite{Schaefer:2008hk}
and in the presence of the Polyakov loop in \cite{Gupta:2009fg} at
$\mu=0$ and in \cite{Tiwari:2013pg} at $\mu\ne0,$ while $\Delta
m^2_{i,ab}$ was first computed in \cite{Chatterjee:2011jd}. We shall
review below the expressions of $\Delta m^2_{i,ab}$ and $\delta
m^2_{i,ab},$ and in Table~\ref{Tab:m2_matrix} we explicitly give their
contributions to the tree-level masses, which are also listed
there. Note that in the $N-S$ basis there are no off-diagonal
contributions to the curvature matrix coming from the fermionic
fluctuations. In the respective mixing sector, the eigenvalues
$m^2_{\eta'/\eta}$ and $m_{f_0^H/f_0^L}^2$ can be computed with the
following formulas:
\begin{align}
  m_{\eta^{\prime}/\eta}^{2} & =\frac{1}{2}\left[ 
    m_{\eta_{N}}^{2} +m_{\eta_{S}}^{2}\pm\sqrt{(m_{\eta_{N}}^{2}-m_{\eta_{S}}^{2})^{2}+4m_{\eta_{NS}}^{4}}\,\right],
  \label{Eq:m_eta_etap}\\
  m_{{f_{0}^{H}}/{f_{0}^{L}}}^{2} & =\frac{1}{2}\left[ 
    m_{\sigma_{N}}^{2}+m_{\sigma_{S}}^{2}\pm\sqrt{(m_{\sigma_{N}}^{2} - m_{\sigma_{S}}^{2})^{2} + 4m_{\sigma_{NS}}^{4}}\,\right].
  \label{Eq:m_f0L_f0H}
\end{align}

The fermionic vacuum contribution to the curvature mass is given by
\begin{align}
\Delta m^2_{i,ab} &= \frac{\partial^2 \Omega_{q\bar  q}^{(0)\textnormal{v}}} {\partial \varphi_{i,a}\partial
  \varphi_{i,b}} \bigg|_\textnormal{min}\nonumber  \\ 
&= -\frac{3}{8\pi^2}\sum_{f=u,d,s}\biggl[\bigg(\frac{3}{2} +
\log\frac{m_f^2}{M_0^2} \bigg)\,{m^2_{f,a}}^{\!\!\!\!\!(i)}{m^2_{f,b}}^{\!\!\!\!\!(i)} \nonumber  \\ 
&+ m_f^2\bigg(\frac{1}{2} + \log\frac{m_f^2}{M_0^2} \bigg) {m^2_{f,ab}}^{\!\!\!\!\!\!\!(i)}\biggr],
\label{eq:DM2i_ab}
\end{align}
where we introduced, as in \cite{Schaefer:2008hk}, shorthands for the
first and second derivatives of the constituent quark mass squared
with respect to the meson fields: ${m^2_{f,a}}^{\!\!\!\!\!(i)}
\equiv \partial m^2_f/\partial \varphi_{i,a}$ and
${m^2_{f,ab}}^{\!\!\!\!\!\!\!(i)} \equiv \partial^2 m^2_f/\partial
\varphi_{i,a} \partial \varphi_{i,b}.$

The correction to the curvature matrix due to the fermionic thermal
fluctuations in the presence of the Polyakov loop reads
\begin{align}
  \delta m^2_{i,{ab}} &= \frac{\partial^2 \Omega_{q\bar q}^{(0)\textnormal{T}}}{\partial \varphi_{i,a} \partial \varphi_{i,b}} \bigg|_\textnormal{min} 
  = 6\sum_{f=u,d,s} \int \frac{d^3 p}{(2\pi)^3} \frac{1}{2E_f(p)} \nonumber\\
  & \times \biggl[\big(f_f^+(p) + f_f^-(p)\big) 
  \biggl( {m^{2}_{f,a b}}^{\!\!\!\!\!\!\!(i)} -
  \frac{{m^{2}_{f,a}}^{\!\!\!\!\!(i)} {m^{2}_{f, b}}^{\!\!\!\!\!(i)}}{2 E_f^2(p)} \biggr)  \nonumber \\
  &+ \big(B_f^+(p) + B_f^-(p)\big) \frac{{m^2_{f,a}}^{\!\!\!\!\!(i)}
    {m^2_{f,b}}^{\!\!\!\!\!(i)}}{2 T E_f(p)}  \biggr],   
\label{Eq:dM2i_ab}
\end{align}
where 
\begin{equation}
  f^\pm_f(p) =\frac{ \Phi^\pm e^{-\beta E_f^\pm(p)} + 2\Phi^\mp e^{-2 \beta E_f^\pm(p)} + e^{-3\beta E_f^\pm(p)} } {g^\pm_f(p)}, 
\label{Eq:mod_FD}
\end{equation}
is the modified Fermi-Dirac distribution functions for quarks
($+$) and antiquarks ($-$) and, following Ref.~\cite{Gupta:2009fg}, we
also introduced $B_f^\pm(p)=3(f_f^\pm(p))^2 - C_f^\pm(p)$ with
\begin{equation}
  C_f^\pm(p) = \frac{\Phi^\pm e^{-\beta E_f^\pm(p)} + 4 \Phi^\mp e^{-2\beta E_f^\pm(p)} +3 e^{-3\beta E_f^\pm(p)}}{g^\pm_f(p)}. 
\label{Eq:C_Phi_barPhi}
\end{equation}

To obtain the mass squares, whose first and second derivatives
appear in Eqs.~\eqref{eq:DM2i_ab} and \eqref{Eq:dM2i_ab}, we have to
find the eigenvalues of the square of the $\gamma_0 \mathcal{M}$
matrix from Eq.~\eqref{Eq:Zqq0}, which is a $12\times 12$ matrix in
the Dirac and flavor space, or, equivalently, of the matrix
$\mathcal{N  N^\dagger},$ where $\mathcal{N}=\sigma_a\lambda_a +
i\pi_a\lambda_a$, which is a $3\times 3$ matrix. An easy way to do the
calculation of a given derivative is to set to zero all the
fluctuating fields not used in the differentiation. The calculation is
straightforward and as noted in \cite{Schaefer:2008hk}, some
cancellations occur in the isospin symmetric case, where the mass
squared of the two light quarks can be combined. The result is given in
the $N-S$ basis in Table~\ref{Tab:qmassd} which, in the case of the
L$\sigma$M, appeared first in the $0-8$ basis in \cite{Schaefer:2008hk}.

\begin{table}
\caption{The first and second derivatives of the quark squared masses with
  respect to the scalar (S) and pseudoscalar (P) meson fields,
  evaluated in the $N-S$ basis at the extremum of the grand
    potential.  All entries of the omitted $ab=$NS rows are
  vanishing. The result holds in the isospin symmetric case and a
  summation over $l\in\{u,d\}$ is understood in the first two columns.
\label{Tab:qmassd}}
\begin{ruledtabular}
\begin{tabular}{cc|cccc}
$i$ & $ab$ & ${m^{2}_{l,a}}^{\!\!\!\!(i)}
{m^{2}_{l,b}}^{\!\!\!(i)}/g^4_F$ &
${m^{2}_{l,ab}}^{\!\!\!\!\!\!(i)}/g^2_F$ &
${m^{2}_{s,a}}^{\!\!\!\!\!(i)} {m^{2}_{s,b}}^{\!\!\!\!(i)}/g^4_F$ &
${m^{2}_{s,ab}}^{\!\!\!\!\!\!\!(i)}/g^2_F$\\
\hline
S & 11 & $\frac{1}{2}\phi_N^2$ & $ 1$ & $0$ & $0$\\
S & 44 & $0$ & $\frac{Z_{K_0^\star}^2\phi_N}{\phi_N - \sqrt{2}
  \phi_S}$ & $0$ & $\frac{-\sqrt{2}Z_{K_0^\star}^2\phi_S}{\phi_N - \sqrt{2}\phi_S}$ \\
S & NN & $\frac{1}{2}\phi_N^2$ & $1$ & $0$ & $0$ \\
S & SS & $0$ & $0$ & $\phi_S^2$ & $1$\\ \hline  
P & 11 & $0$ & $Z_\pi^2$ & $0$ & $0$ \\
P & 44 & $0$ & $\frac{Z_K^2\phi_N}{\phi_N +\sqrt{2}\phi_S}$ & $0$ & $\frac{\sqrt{2}Z_K^2\phi_S}{\phi_N + \sqrt{2} \phi_S}$ \\
P & NN & $0$ & $Z_{\eta_N}^2$ & $0$ & $0$\\
P & SS & $0$ & $0$ & $0$ & $Z_{\eta_S}^2$
\end{tabular}
\end{ruledtabular}
\end{table}

For $\bar\Phi=\Phi=1,$ the distribution functions $f^\pm_f(p)$ goes
over into the usual Fermi-Dirac distributions for quarks
and antiquarks,
$f^\pm_f(p)\to f^\pm_{f,\textnormal{FD}}(p)=1/(e^{\beta(E_f(p)\mp\mu_f)}+1).$
In this limit, which is expected to be reached at high temperature,
$B_f^\pm(p)\to -f^\pm_{f,\textnormal{FD}}(p)(1-f^\pm_{f,\textnormal{FD}}(p)),$
and one recovers the expression of Ref.~\cite{Schaefer:2008hk} for the
curvature mass, obtained in the linear sigma model without the
inclusion of the Polyakov loop. When $\bar\Phi=\Phi=0,$ which is 
reached for vanishing temperature, the so-called 
``statistical confinement'' occurs, as 
$f^\pm_f(p)\to 1/(e^{\beta(3E_f(p)\mp \mu_f)}+1),$
which means that at small temperature three quark states, that is
excitations with zero triality, represent the effective degrees of freedom
\cite{Fukushima:2003fw}.

\subsection{Field equations}
\label{subsec:field_eqns}

Up to this point we were quite formal in dealing with the consequence
of the quark's propagation on a constant gluon background field in the
temporal direction.  Now we have to face the situation that, since
$\Phi$ and $\bar\Phi$ are complex, the grand potential we arrived at
is, in fact, a complex function of complex variables. It is easy to
see that $\Omega_{\bar q q}^{(0)\textnormal{T}}(T,\mu_q)$ in
\eqref{Eq:fermion_thermal} has an imaginary part for $\mu_q\ne0$,
which is the manifestation of the sign problem in the present context,
and the question is how to extract physical information from the grand
potential (see also the discussion in \cite{Mintz:2012mz}).  In the
mean-field approximation of Ref.~\cite{Ratti:2005jh} (see also \cite
{Rossner:2007ik}) the traced Polyakov loops $\Phi$ and $\bar\Phi$
introduced in \eqref{Eq:Phi_barPhi} are replaced by their thermal
expectation values $\langle\Phi\rangle$ and $\langle\bar\Phi\rangle$,
that is by the Polyakov loop variables, which at $\mu_B\ne0$ are
treated as two real and independent quantities (at $\mu_B=0$ they are
equal). Adopting this approach and using for simplicity the notation
$\Phi$ and $\bar\Phi$ for the Polyakov loop variable, it is understood
that from now on in Eqs.~\eqref{Eq:g_Phi_barPhi}, \eqref{Eq:mod_FD},
and \eqref{Eq:C_Phi_barPhi} the fields $\Phi$ and $\bar\Phi$ are real
and independent. In this approach the grand potential $\Omega$ is real
and the physical point (extremum of $\Omega$) is a saddle point.
Working with real Polyakov loop variables $\Phi$ and $\bar\Phi$ seems
to be supported by the study performed in the massive extension of the
Landau-DeWitt gauge, where the self-consistent gauge fixing condition
imposes constraints on the background gauge fields $\bar A^3$ and
$\bar A^8$ [which correspond to $\phi_3$ and $\phi_8$ of
\eqref{eq:G4_form}].  As the study in \cite{Reinosa:2015oua} reveals,
for real values of $\mu_B$ the constraints are obeyed by real $\bar
A^3$ and imaginary $\bar A^8$ gauge fields, and these field
configurations correspond to real and independent Polyakov loop
variables $\Phi$ and $\bar\Phi.$\footnote{We thank Urko Reinosa for
  explaining to us the relevance of his works in the present context
  and for sharing with us the ideas and subtleties related to the
  construction of a physically meaningful potential.} We mention that
in some cases, another approach is preferred in the PL$\sigma$M, in
which the imaginary part of the potential is neglected
\cite{Mintz:2012mz}.  In this case the physical point is a minimum,
which makes possible the investigation of the nucleation occurring
during a first order phase transition, but has the drawback that the
difference between the expectation values of the traced Polyakov loop
and its conjugate vanishes at $\mu_B\ne0$.

In view of the above discussion, the field equations, which determine
the dependence on $T$ and $\mu_B=3\mu_q$ of the chiral condensates
$\phi_N$ and $\phi_S$ and the Polyakov loop variables $\Phi$ and
$\bar\Phi,$ are obtained by extremizing the grand potential,
\begin{equation} 
\frac{\partial\Omega_\textnormal{H}}{\partial \phi_N} =
\frac{\partial\Omega_\textnormal{H}}{\partial \phi_S} =
\frac{\partial\Omega_\textnormal{H}}{\partial \Phi} =
\frac{\partial\Omega_\textnormal{H}}{\partial \bar\Phi} = 0.
\end{equation}
In our hybrid approach we include in the field equations only the
vacuum and thermal fluctuations of the constituent quarks and leave
out the corresponding mesonic fluctuations. In this case, the explicit
field equations read
\begin{widetext}
\begin{subequations}
\begin{align}
  &-\frac{d}{d \Phi}\left( \frac{U(\Phi,\bar\Phi)}{T^4}\right) + \frac{6}{T^3}\sum_{f=u,d,s} 
  \int \frac{d^3p}{(2\pi)^3} \left(\frac{e^{-\beta E_f^-(p)}}{g_f^-(p)}
    + \frac{e^{-2\beta E_f^+(p)}}{g_f^+(p)} \right) = 0, \label{eq_Phi}\\
  &-\frac{d}{d \bar\Phi}\left( \frac{U(\Phi,\bar\Phi)}{T^4}\right) + \frac{6}{T^3}\sum_{f=u,d,s}
  \int \frac{d^3p}{(2\pi)^3} \left(\frac{e^{-\beta E_f^+(p)}}{g_f^+(p)}
    + \frac{e^{-2\beta E_f^-(p)}}{g_f^-(p)} \right) = 0, \label{eq_Phibar}\\
  &m_0^2 \phi_N + \left(\lambda_1 + \frac{1}{2} \lambda_2 \right)
  \phi_N^3 + \lambda_1 \phi_N \phi_S^2 -\frac{1}{\sqrt{2}} c_1\phi_N\phi_S - h_{0N}
  +\frac{3}{2}g_F\left(\langle \bar q_u q_u\rangle_{_{T}} + \langle\bar q_d q_d\rangle_{_{T}} \right) = 0,\label{eq_phiN}\\
  &m_0^2 \phi_S + \left(\lambda_1 + \lambda_2 \right) \phi_S^3 +
  \lambda_1 \phi_N^2 \phi_S -\frac{\sqrt{2}}{4}c_1\phi_N^2 - h_{0S} 
  +\frac{3}{\sqrt{2}}g_F \langle \bar q_s q_s\rangle_{_{T}} = 0,\label{eq_phiS}
\end{align}
\end{subequations}
\end{widetext}
where $U(\Phi,\bar\Phi)$ is the Polyakov loop potential
\eqref{eq:Ulog} and, by matching the renormalization of the effective
potential done in Sec.~\ref{Subsec:ren}, we defined the renormalized
expectation value\footnote{It is worth noting that the expectation
  value $\langle \bar q_f q_f\rangle$ is calculated within the
  framework of the present model containing constituent quarks and it
  is not directly related to $\langle \bar q q\rangle$ appearing in
  the QCD.} as
\begin{equation}
  \langle \bar q_f q_f\rangle_{_{T}} = 4 m_f
  \left[ -\frac{m_f^2}{16\pi^2}\left(\frac{1}{2} +
      \ln\frac{m_f^2}{M_0^2} \right) + T_f\right],
\end{equation}
with the thermal part of the fermion tadpole integral given by
\begin{equation}
  T_f = \int \frac{d^3p}{(2\pi)^3}\frac{1}{2E_f(p)}\big(f^-_f(p) +
  f^{+}_{f}(p)\big) \,.
\label{Eq:tad_th}
\end{equation}

\section{Determination of the model parameters}
\label{Sec:parametrization}

There are altogether $16$ unknown parameters, $15$ parameters found in
the Lagrangian given in \eref{Eq:Lagr} and the renormalization scale
$M_0$ (see \secref{Subsec:ren}). For the renormalization scale we
choose three different initial values, namely $M_0=0.3, 0.9, 1.5$~GeV
and run the parametrization for them. After finding a good
solution---which includes a particular $M_0$ value---we take it as
an initial condition and minimize for $M_0$ around that solution. From
the remaining $15$ Lagrangian parameters $\delta_N$ can be
incorporated (without loss of generality) into $m_1$---the bare
(axial)vector mass, while the external fields $h_{0N}$ and $h_{0S}$
are replaced by the scalar condensates $\phi_N$ and $\phi_S$ with the
help of the field equations \eqref{eq_phiN} and \eqref{eq_phiS} at
zero temperature. Consequently, there are $14$ parameters left to be
determined, which are the following: the bare (pseudo)scalar mass $m_0$; the
(pseudo)scalar self-couplings $\lambda_1$ and $\lambda_2$; the
$U_{A}(1)$ anomaly coupling $c_1$; the bare (axial)vector mass $m_1$;
the (axial)vector--(pseudo)scalar couplings $h_1$, $h_2$ and $h_3$;
the external field $\delta_S$ which explicitly breaks the chiral
symmetry in the (axial)vector sector; the scalar condensates $\phi_N$
and $\phi_S$; the Yukawa coupling $g_F$; and two (axial)vector
couplings $g_1$ and $g_2$.

In the parametrization procedure we use alongside $29$ vacuum
quantities, that is 15 masses, 12 tree-level decay widths, 2 Partially
Conserved Axialvector Current (PCAC) relations
$f_{\pi}=\phi_N/Z_{\pi}$ and
$f_{K}=(\phi_N+\sqrt{2}\phi_S)/(2Z_{K}),$, and also the pseudocritical
temperature $T_c$ (see the next paragraph for explanation). The masses
used are the following: the curvature masses of the pseudoscalars
$m_{\pi}$, $m_{K}$, $m_{\eta}$, $m_{\eta^{\prime}}$ and the scalars
$m_{a_0}$, $m_{K_0^{\star}}$, $m_{f_0^L}$, $m_{f_0^H}$ listed in
\tabref{Tab:m2_matrix}, where the fermionic corrections to the
tree-level masses are also given [see also Eqs.~\eqref{Eq:m_eta_etap}
and \eqref{Eq:m_f0L_f0H}]; the tree-level masses of the vector mesons
$m_{\rho}=m_{\omega}$, $m_{K^{\star}}$, $m_{\Phi}$, the axial vector
mesons $m_{a_1}=m_{f_1^L}$, $m_{f_1^H}$ to be found in
\cite{elsm_2013}; the tree-level constituent quark masses $m_{u,d}$,
$m_s$ given in \eqref{Eq:uds_masses}.  \footnote{Note that the
  relations $m_{\rho}=m_{\omega}$ and $m_{a_1}=m_{f_1^L}$ hold at tree
  level in our model and that we do not use $m_{K_1}$. For the latter
  see the discussion in \cite{elsm_2013}.} The decay widths used are
the vector decays $\Gamma_{\rho\to\pi\pi}$, $\Gamma_{K^{\star}\to
  K\pi}$, $\Gamma_{\Phi\to K K}$, the axial vector decays
$\Gamma_{a_1\to\rho\pi}$, $\Gamma_{a_1\to\pi\gamma}$, $\Gamma_{f_1\to
  K^{\star}K}$, and the scalar decays $\Gamma_{a_0}$,
$\Gamma_{K_0^{\star}}$, $\Gamma_{f_0^L\to\pi\pi}$, $\Gamma_{f_0^L\to K
  K}$, $\Gamma_{f_0^H\to\pi\pi}$, $\Gamma_{f_0^H\to K K}$ given in
\cite{elsm_2013} and Appendix~\ref{App:decays}. The value of the
masses and decay constants are compared with the corresponding
experimental value taken from the PDG \cite{PDG} through the $\chi^2$
minimization method of Ref.~\cite{MINUIT} similarly as in
\cite{elsm_2013}, but with some important differences listed below.
One such difference, mentioned already and detailed more latter, is
the inclusion of the pseudocritical temperature $T_c$ in the
minimization process.  We take the mean value given in the PDG (in
case of charged particles, the neutral and charged masses are
averaged) and for the error we allow for a $20\%$ variation with
respect to the PDG value for the masses and decay widths of the scalar
sector, $10\%$ for the constituent quark masses and $5\%$ for all the
other quantities.  We use this large error in case of the scalars
mainly because they mix with each other and our fields do not
correspond to pure physical particles, while in case of the
constituent quarks, their dynamically generated mass depends on how it
is defined and calculated.  All other errors of the masses and decay
widths of the pseudoscalars, vectors and axial vectors are much
smaller experimentally; however, we used $5\%$ for them due to model
limitations and approximations (e.g., isospin symmetry).  All the data
used for the parametrization are listed in
Appendix~\ref{App:param_data}.

Compared to \cite{elsm_2013}, the modifications in the
parametrization of the model are the following.
\begin{enumerate}[(i)]
\item Since here we use a different anomaly term [see
  \eqref{Eq:Lagr}], the terms proportional to $c_1$ are different in
  the expressions of the tree-level pseudo(scalar) masses and the
  scalar decay widths, which are listed explicitly in the first column
  of Table~\ref{Tab:m2_matrix} and in \apref{App:decays},
  respectively. The expressions of the (axial)vector meson masses and
  decay widths are unchanged.
\item A small modification in the present case is that for the
  $a_0(980)$ particle we fit to the value of the total width found in
  \cite{PDG}, instead of fitting to the value of the two amplitudes
  $|\mathcal{M}_{a_0(980)\to KK}|$ and $|\mathcal{M}_{a_0(980)\to
    \eta\pi}|$ found in \cite{giacosa_2010}.
\item We now include the $f_0$ masses and decay widths into the global
  fit, as opposed to \cite{elsm_2013}, where we first did a global fit
  without using the properties of the $f_0$ mesons and only after that
  we analyzed the consequences of the fit on the $f_0$'s.
\item We consider here the effects of the fermion vacuum fluctuations,
  case in which the expression of the (pseudo)scalar masses are
  modified, as shown in the second column of
  Table~\ref{Tab:m2_matrix}.
\item Working in the isospin symmetric limit, we use now the two
  additional tree-level equations for the constituent quark masses
  given in \eqref{Eq:uds_masses}. Their explicit expression contains
  the Yukawa coupling $g_F$ and the values $m_{u,d} = 308$~MeV and
  $m_{s} = 483$~MeV were used for the fit. These values are obtained
  from a nonrelativistic mass formula for the light mesons in which
  spin-spin interaction is taken into account, as presented in Chap.~5.5
  of Ref.~\cite{Griffiths:2008zz}.
\end{enumerate}

As was discussed in \cite{elsm_2013}, the scalar sector below $2$~GeV
contains more physical particles than states in one $q\bar q$ nonet
(consisting of $a_0$, $K_0^{\star}$, $f_0^L$, $f_0^H$), since in nature
there are two $a_0$, two $K_0^{\star}$, and five $f_0$ particles in the
considered energy range. These particles are the $a_0(980)$ and
$a_0(1450)$, which will be denoted by $a_0^1$ and $a_0^2$; the
$K_0^{\star}(800)$ and $K_0^{\star}(1430)$, which will be denoted by
$K_0^{\star\, 1}$ and $K_0^{\star\, 2}$; and the $f_0(500)$ [previously called
as $\sigma$ or $f_0(600)$], $f_0(980)$, $f_0(1370)$, $f_0(1500)$, and
$f_0(1710)$, which will be denoted by $f_0^{1}, \dots, f_0^{5}$,
respectively.  Consequently, there are 40 possibilities to assign the
existing scalar physical particles to the corresponding scalar nonet states.

Since compared to \cite{elsm_2013} our parametrization considerably
changed due to the inclusion of the $f_0$ masses, decay widths, and
fermionic vacuum fluctuations, we reran the fitting procedure for all
40 cases and for every $M_0$ value mentioned earlier and retained
only those solutions of the $\chi^2$ minimization, which gave the
lowest $\chi^2$ values. However, by using only zero temperature
quantities (PCAC relations, masses and decay widths) in the
parametrization we would end up with lots of possible solutions with
very close $\chi^2$ values, which could produce various, even
physically unacceptable, thermodynamical behaviors.  More specifically,
the $T_c$ pseudocritical temperature at zero baryon chemical
potential, which should be around $150$~MeV,\footnote{Continuum
  extrapolated lattice results give $T_c=151$~MeV from the peak of the
  chiral susceptibility \cite{Aoki:2006br} and $T_c=157$~MeV if the
  inflection point of the subtracted chiral condensate is used
  \cite{Borsanyi:2010bp}.}  can reach very high values ($\gtrapprox
350$~MeV) in case of some solutions.  Thus we chose to include the
physical value of $T_c$ in the parametrization with a $10\%$ error.
Additionally, we only considered solutions that had $T_c < 180$~MeV.
For the determination of the $T_c$ we solved the four coupled field
equations Eqs.~\eqref{eq_Phi}\,-\,\eqref{eq_phiS} at $\mu_B=0$ and
defined $T_c$ as the temperature for which the value of the so-called
subtracted chiral condensate is $0.5$.  This quantity, defined in
\cite{Cheng:2007jq} as
\begin{equation}
  \Delta_{l,s}(T) = \frac{(\phi_N - \frac{h_{0N}}{h_{0S}}
    \phi_S)\vert_{T}}{(\phi_N - \frac{h_{0N}}{h_{0S}} \phi_S)
    \vert_{T=0}},
\label{Eq:sub_chi_cond}  
\end{equation}
can be measured on the lattice, and it takes values between $0$ and $1.$

The $\chi^2$ and $\chi^2_{\text{red}}\equiv \chi^2 /N_\text{dof}$
values\footnote{The number of the degrees of freedom, $N_\text{dof}$
  is the difference between the number of fitted quantities (30) and
  the number of fitting parameters (14).  Note that $M_0$ is kept
  fixed.} for the first ten best solutions are shown in
Table~\ref{Tab:chi2_best_10} along with the corresponding particle
assignments. Interestingly, in the case of the ten best solutions the
value of $M_0$ was always $0.3$~GeV (from the three possibilities
$0.3, 0.9$, and $1.5$~GeV). In these solutions we used the logarithmic
Polyakov loop potential \eqref{eq:Ulog} with
$T_0=182$~MeV. Considering that we would like to carry out the
thermodynamical analysis with one particular set of parameters (which
means one particular assignment of the scalar states), we could simply
choose the first one. However, since the first couple of solutions are
not very far from each other in $\chi^2$ values, it is better if we
take a closer look at the details of the fits and see how well they
describe the spectrum physically. The detailed fit results of the
first two best solutions are shown in
\tabref{Tab:observ_particle_setup} of Appendix~\ref{App:param_data}
together with the result taken from \cite{elsm_2013}.  In the case of
the two best solutions the majority of the $30$ physical quantities
listed there are in good agreement with the experimental
values. However, there are some quantities that are not well
described, like the mass of $a_1$, which we find in any current fit
smaller than its experimental value, and which consequently result in
too small values for the $a_1$ decays as well. Considering the first
assignment $a_0^{1} K_0^{\star\,1} f_0^{1} f_0^{2}$ we cannot see any
inconsistency; on the other hand, in case of the second assignment
(right ``Fit'' column), $a_0^{1} K_0^{\star\,1} f_0^{1} f_0^{3}$, the
$f_0^H$ should correspond to $f_0(1370)$, while the fitted values of
its mass and $\Gamma_{f_0^H\rightarrow KK}$ decay---which are
$802.4$~MeV and $0$~MeV, respectively---are much closer to the data
of $f_0(980)(\equiv f_0^2)$. Though its other decay turns out to be
$\Gamma_{f_0^H\rightarrow\pi\pi}=249.5$~MeV, this value indeed
belongs to $f_0(1370)$. This means that this assignment can be
excluded even on the grounds of physical inconsistency. With the same
argument all elements of the list in Table~\ref{Tab:chi2_best_10} can
be excluded except one, which is indeed the best solution with
assignment $a_0^{1} K_0^{\star\,1} f_0^{1} f_0^{2}$ (middle ``Fit''
column). Thus we choose the parameter set belonging to the $a_0^{1}
K_0^{\star\,1} f_0^{1} f_0^{2}$ assignment for the thermodynamical
investigations of the next section and minimize for $M_0$, which
reduces the $\chi^2$ slightly to $18.53$. The corresponding values of
the parameters are given in \tabref{Tab:param}. Using
Eqs.~\eqref{eq_phiN} and \eqref{eq_phiS} at $T = 0$ one obtains for
the value of the external fields $h_{0N}=(108.488\ \text{MeV})^3$ and
$h_{0S}=(287.832\ \text{MeV})^3.$

\begin{table}[t]
 \caption{$\chi^2$ and $\chi^2_{\text{red}}=\chi^2/N_\text{dof}$ values
   ($N_\text{dof}=16$, because $M_0$ is kept fixed) for the first ten best
   solutions of the fit together with the corresponding physical scalar
   meson particle assignment.  See the text for the meaning of the
   superscript in the particle assignment.  \label{Tab:chi2_best_10}}
\begin{ruledtabular}
\begin{tabular}{ccc}
Particle assignment & $\chi^2$ & $\chi^2_{\text{red}}$\\
\hline
$a_0^{1} K_0^{\star\, 1} f_0^{1} f_0^{2}$ & 18.57 &  1.16 \\ 
$a_0^{1} K_0^{\star\, 1} f_0^{1} f_0^{3}$ & 21.38 &  1.34 \\ 
$a_0^{1} K_0^{\star\, 2} f_0^{1} f_0^{3}$ & 27.80 &  1.74 \\ 
$a_0^{1} K_0^{\star\, 2} f_0^{1} f_0^{2}$ & 28.42 &  1.77 \\ 
$a_0^{1} K_0^{\star\, 1} f_0^{2} f_0^{3}$ & 29.37 &  1.83 \\ 
$a_0^{2} K_0^{\star\, 1} f_0^{1} f_0^{2}$ & 31.65 &  1.98 \\ 
$a_0^{2} K_0^{\star\, 1} f_0^{1} f_0^{3}$ & 33.41 &  2.09 \\ 
$a_0^{1} K_0^{\star\, 2} f_0^{2} f_0^{3}$ & 35.99 &  2.25 \\ 
$a_0^{1} K_0^{\star\, 1} f_0^{1} f_0^{5}$ & 38.87 &  2.43 \\ 
$a_0^{2} K_0^{\star\, 1} f_0^{2} f_0^{3}$ & 41.54 &  2.60 \\ 
\end{tabular}
\end{ruledtabular}
\end{table}

It is worth noting that according to \cite{elsm_2013} without fitting
the $f_0^{L/H}$ mesons the best solution is the combination $a_0^2
K_0^{\star\, 2}$, and we argued that with that solution the most
favorable $f_0^{L/H}$ assignment is the $f_0^{3/5}$. For a general
investigation the procedure followed in \cite{elsm_2013} is the right
strategy, since the physically observed $f_0$ states are probably
mixtures of elementary diquark, tetraquark and glueball states (from
which the latter ones are not included in the present model);
therefore, our $f_0^{L/H}$ states cannot be identified directly with
any of the $f_0^i$ states. Since we could not quantify that mixing, we
left out the $f_0$'s from the fit. However, in this study, the
thermodynamical properties of the system depend on $f_0^{L}$ very
strongly, and thus we had to identify it with one of the $f_0^i$ states
and include it in the parametrization.

\begin{table}[t]
\caption{Parameter values in the case of the  $a_0^{1} K_0^{\star\,1}
  f_0^{1} f_0^{2}$ particle assignment obtained using the logarithmic 
Polyakov loop potential \eqref{eq:Ulog} with $T_0=182$~MeV. \label{Tab:param}}
\centering
\begin{tabular}[c]{|c|c||c|c|}\hline
Parameter & Value & Parameter & Value \\\hline\hline
$\phi_{N}$ [GeV]& $0.1411$ & $g_{1}$ & $5.6156$\\\hline
$\phi_{S}$ [GeV]& $0.1416$ & $g_{2}$ & $3.0467$\\\hline
$m_{0}^2$ [GeV$^2$] & $2.3925\e{-4}$ & $h_{1}$ & $27.4617$\\\hline
$m_{1}^2$ [GeV$^2$] & $6.3298\e{-8}$ & $h_{2}$ & $4.2281$\\\hline
$\lambda_{1}$ & $-1.6738$ & $h_{3}$ & $5.9839$\\\hline
$\lambda_{2}$ & $23.5078$ & $g_{F}$ & $4.5708$\\\hline
$c_{1}$ [GeV]& $1.3086$ & $M_{0}$ [GeV]& $0.3511$\\\hline
$\delta_{S}$ [GeV$^2$] & $0.1133$ & \multicolumn{1}{c}{} & \multicolumn{1}{c}{} \\\cline{1-2}
\end{tabular}
\end{table}

\section{Results}
\label{Sec:result}

In this section we present the dependence on the temperature and chemical
potential of various physical quantities determined with the best set of
parameters found with our parametrization procedure. We compare the
variation of the condensates and that of the pressure and the quantities
derived from it, like the energy density, the interaction measure, and the
speed of sound, with recent continuum extrapolated lattice results.  In
doing so, we vary the parameter $T_c^\text{glue}$ of the improved Polyakov
loop potential \eqref{Eq:U_glue} in the range of $[182, 270]$~MeV, in an
attempt to see whether the lattice result could restrict its value. 
Changing $T_c^\text{glue}$ affects the value of $T_c,$ but it does not
affect the vacuum value of the quantities used for parametrization. We also
study the $\mu_B-T$ phase diagram and the existence of the CEP.

\subsection{Temperature variation of condensates and meson masses at
  $\mu_B=0$}
\label{SubSec:cond_mass}

\begin{figure*}[!t]
\centerline{\includegraphics[width=0.9\textwidth]{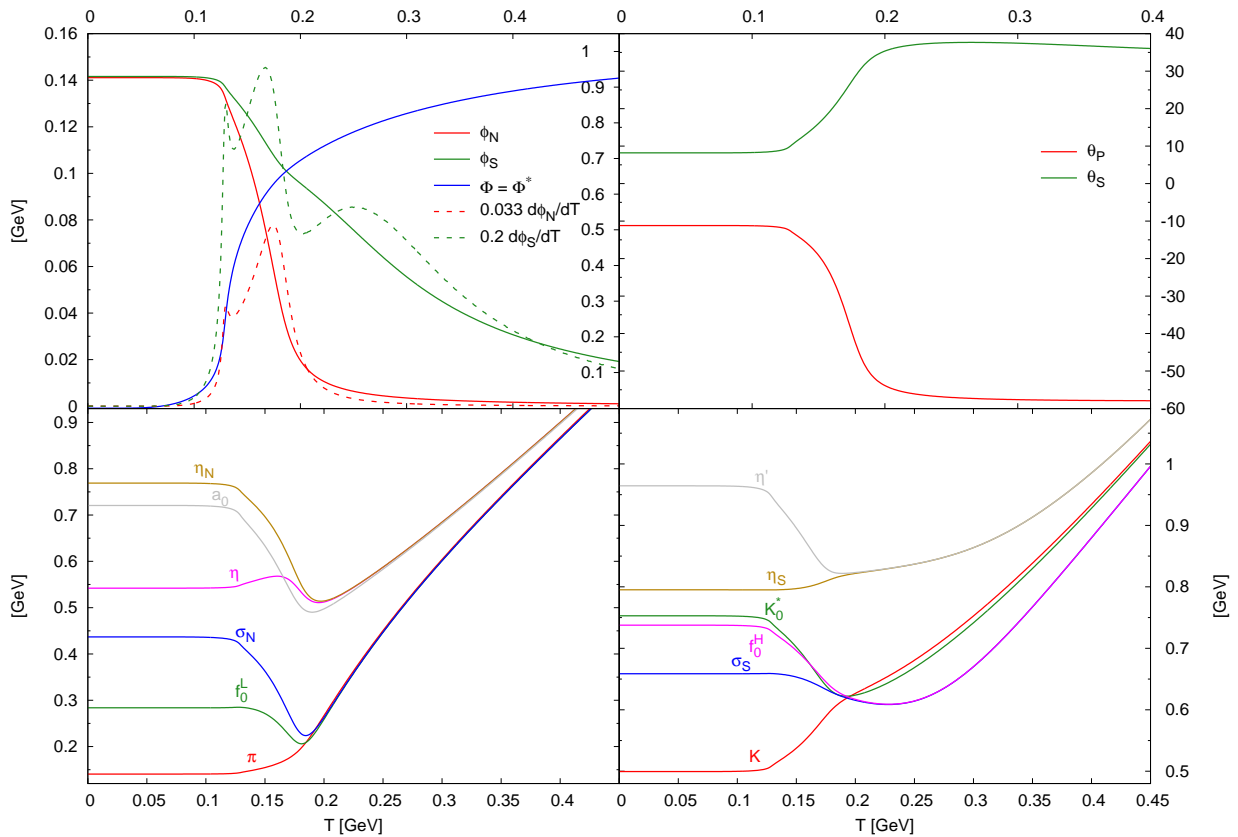}}
\caption{Temperature dependence of various quantities obtained at
  $\mu_B=0$ from the improved Polyakov loop potential $U_\text{glue}$
  \eqref{Eq:U_glue} proposed in \cite{Haas:2013qwp} (here we used
  $T_c^\text{glue}=182$~MeV). Top left: nonstrange and strange chiral
  condensates along with their temperature derivative and the Polyakov
  loop expectation value; top right: scalar and pseudoscalar mixing
  angles; and bottom left and right: scalar and pseudoscalar curvature
  masses arranged according to chiral partners $(\pi, f_0^L)$, $(\eta,
  a_0)$ and $(K,K_0^\star),$ $(\eta',f_0^H),$ respectively.  We also
  show the masses of the strange and nonstrange components of the two
  mixing sectors.}
\label{Fig:cond_mass}
\end{figure*}

In Fig.~\ref{Fig:cond_mass} we study at $\mu_B=0$ the temperature variation
of the nonstrange and strange chiral condensates, Polyakov loop expectation
value, scalar and pseudoscalar curvature masses and the corresponding mixing
angles.  These results are obtained using the improved Polyakov loop
potential $U_\text{glue}$ of Eq.~\eqref{Eq:U_glue} with
$T_c^\text{glue}=182$~MeV, \footnote{For other values of $T_c^\text{glue}$
in the $[182, 270]$~MeV interval the curves show a very similar behavior.}
as the value of the critical glue temperature.  We see that the chiral
condensates stay close to their vacuum values up to some quite high value of
the temperature of order 100~MeV. This is the usual manifestation of the
so-called ``Polyakov cooling mechanism'' \cite{Megias:2004hj} already
observed in \cite{Oleszczuk:1992yg}, namely, that when the Polyakov loop is
coupled to chiral quarks, any quark observable at small temperature (deep in
the hadronic phase) takes a value obtained in the theory without the
Polyakov loop at a lower temperature, of the order $T/N_c$. When the values
of the condensates start to drop, a bumpy behavior can be observed in both
the strange and nonstrange condensates.  This behavior, clearly shown by
the temperature derivative of the condensates, is reflected by the
temperature evolution of the masses.

Next, let us consider the restoration of chiral symmetry from the
parity doubling perspective \cite{Costa:2005cz}, that is, by checking
the mass degeneracy of a scalar meson with its opposite-parity
partner.  We see in Fig.~\ref{Fig:cond_mass} that in the nonstrange
sector the $SU(2)$ chiral partners $(\pi, f_0^L)$ and $(\eta, a_0)$
become degenerate at $T\simeq 190$~MeV, which is slightly above the
inflection point ($T_c=172$~MeV) of the nonstrange condensate
$\phi_N(T)$ and subtracted chiral condensate $\Delta_{l,s}(T)$.  In
the strange sector the chiral symmetry is restored at a much higher
temperature, as there is a temperature range of around $200$~MeV where
the masses of the chiral partners $(K,K_0^\star)$ are close, but they
only become degenerate above $T\simeq450$~MeV.  The masses of the
$\eta'$ and $f_0^H$ approach each other, but they never become
degenerate.  This is the consequence of the fact that our anomaly
parameter $c_1$ is temperature independent, and therefore the $U(1)_A$
symmetry is not restored in the explored temperature region.  The
nonrestoration of the $U(1)_A$ symmetry is visible also in the
nonstrange sector, where the axial partners $(\pi,a_0)$ and
$(\eta,f_0^L)$ do not become degenerate. We refer to the literature
for the case when a temperature-dependent anomaly parameter is
considered by using lattice results for the topological
susceptibility.  Typically, following Ref.~\cite{Kunihiro:1989my}, an
anomaly parameter which decreases exponentially with the temperature
or density is considered, which results in a faster restoration of the
chiral symmetry and an effective restoration of the $U(1)_A$ symmetry
\cite{Costa:2005cz,Ruivo:2012xt}.

We turn now to the scalar and pseudoscalar mixing angles in relation
with the masses of the $f_0^L-f_0^H$ and $\eta-\eta'$ complexes.  The
big difference compared to previous results obtained by computing the
grand potential with the same approximation we use here, but without
the inclusion of the (axial)vector mesons, is that in our case, for
temperatures below $0.9$~GeV, one has $m_{f_0^L}\le
m_{\sigma_N}<m_{\sigma_S}\le m_{f_0^H}$ and similarly $m_\eta\le
m_{\eta_N}<m_{\eta_S}\le m_{\eta'}$ in contrast to previous studies,
where $m_{\eta_N}>m_{\eta_S}$ and $m_{\sigma_N}>m_{\sigma_S}$ (see,
e.g., \cite{Schaefer:2008hk}). The temperature evolution of both
mixing angles is such that the situation of ideal flavor mixing is
achieved at temperatures which are $2-3$ times larger than $T_c$; that
is, $f_0^L$ and $\eta$ mesons are pure nonstrange $\bar q q$ states,
while $f_0^H$ and $\eta'$ are pure strange ones.

\begin{figure*}[!t]
  \includegraphics[width=0.485\textwidth]{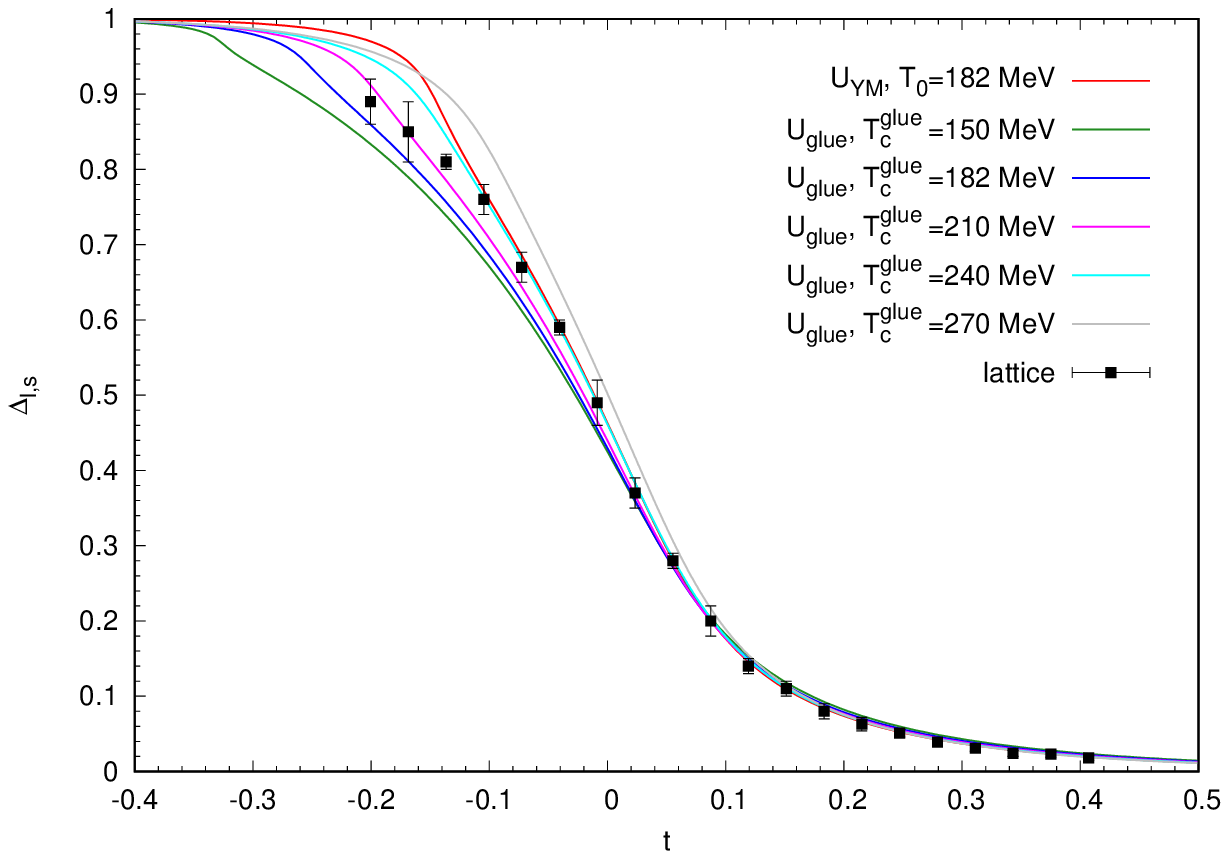}
  \includegraphics[width=0.485\textwidth]{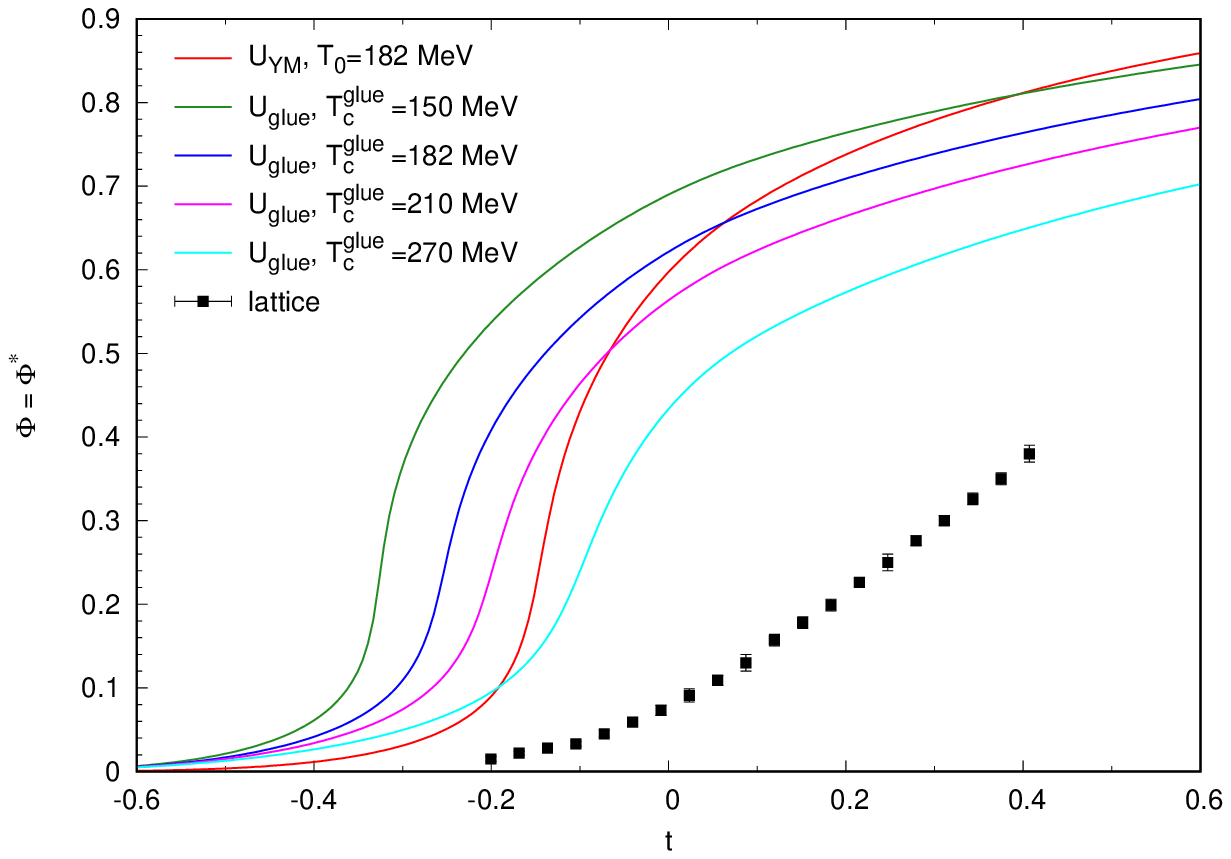}
\caption{The subtracted chiral condensate $\Delta_{l,s}$ given in
  \eqref{Eq:sub_chi_cond} (left panel) and the Polyakov-loop
  expectation values (right panel) determined at $\mu_B=0$ as a
  function of the reduced temperature $t=T/T_c-1$ for different
  parametrizations of the Polyakov loop potential. We compare to the
  continuum extrapolated lattice result of the Wuppertal-Budapest
  Collaboration \cite{Borsanyi:2010bp}.}
\label{Fig:cond_red_temp}
\end{figure*}

Now we look more closely at the thermal evolution of the subtracted chiral
condensate $\Delta_{l,s}$ given in \eqref{Eq:sub_chi_cond} and investigate,
as in Ref.~\cite{Haas:2013qwp}, whether by comparing to the lattice result
it is possible to restrict the values of some parameters of the improved
Polyakov loop potential.  We have already seen that our pseudocritical
temperature is higher than the continuum extrapolated lattice result;
therefore, we plot $\Delta_{l,s}$ as a function of the reduced temperature
$t=T/T_c-1$. To be able to compare with the lattice results of
Ref.~\cite{Borsanyi:2010bp}, we have to assure that we use the same
definition for the pseudocritical temperature.  During the parametrization
we used, as a reasonable and numerically easy to implement approximation for
$T_c,$ the value of the temperature where $\Delta_{l,s}=0.5.$ Now we define
$T_c$ as the inflection point of $\Delta_{l,s}$ obtained by fitting
$f(T)=a+b\arctan(c(\pi - d\,T))$ to our and the lattice
results.\footnote{This procedure, used in \cite{Marko:2013lxa} in the context
of the $O(N)$ model, is accurate in the present context only if the
temperature is restricted to a narrow range around the inflection point.}
Then, the value of the pseudocritical temperature is given by $T_c=\pi/d$. 
Fitting the above function to the lattice data in the range
$T\in[145,165]$~MeV we obtain $T_c=156.35$~MeV, which is compatible with the
reported value $157(3)(3).$ In our case, regarding the logarithmic Polyakov
loop potential with $T_0=182$~MeV we obtain $T_c=172$~MeV, and with the
improved Polyakov loop potential we get $T_c\in(168,189)$~MeV, depending on
the $T_c^\text{glue}$ value used.

\begin{figure*}[!t]
  \includegraphics[width=0.485\textwidth]{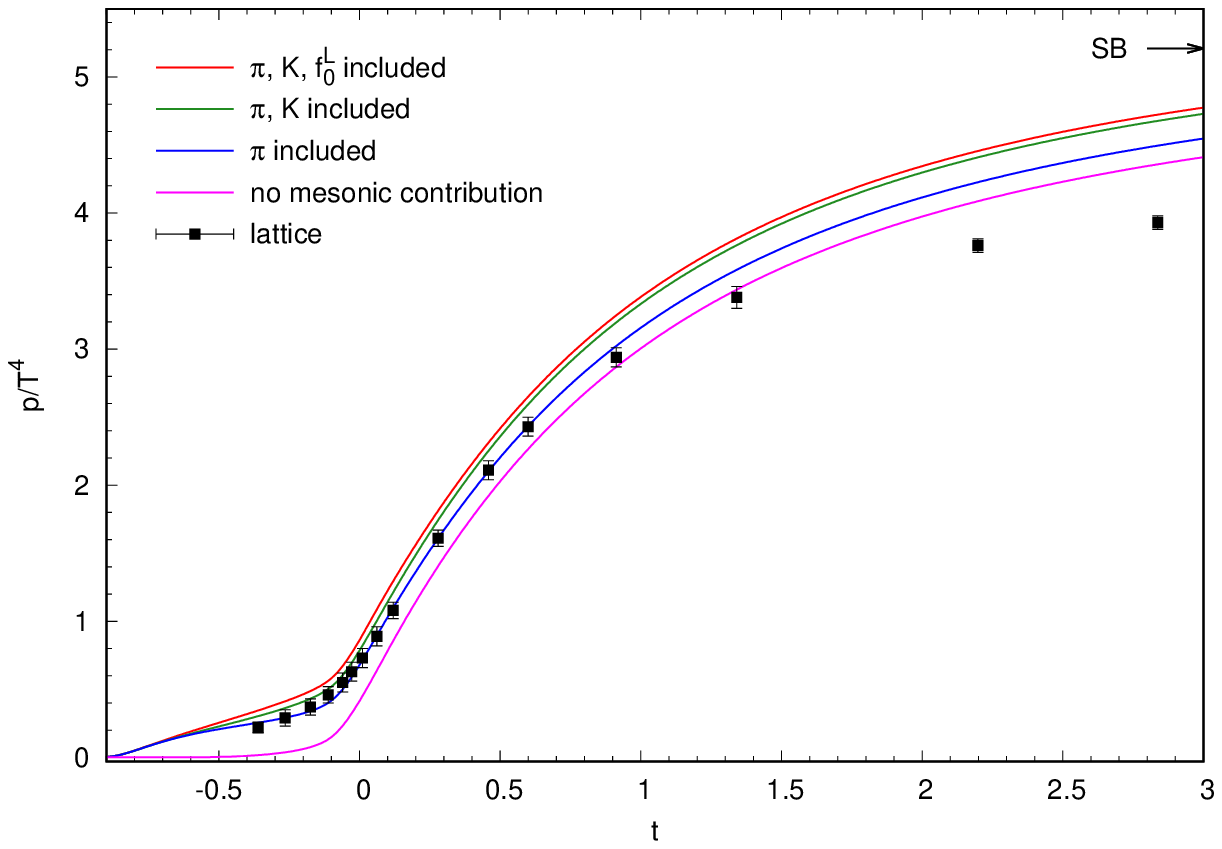}
  \includegraphics[width=0.485\textwidth]{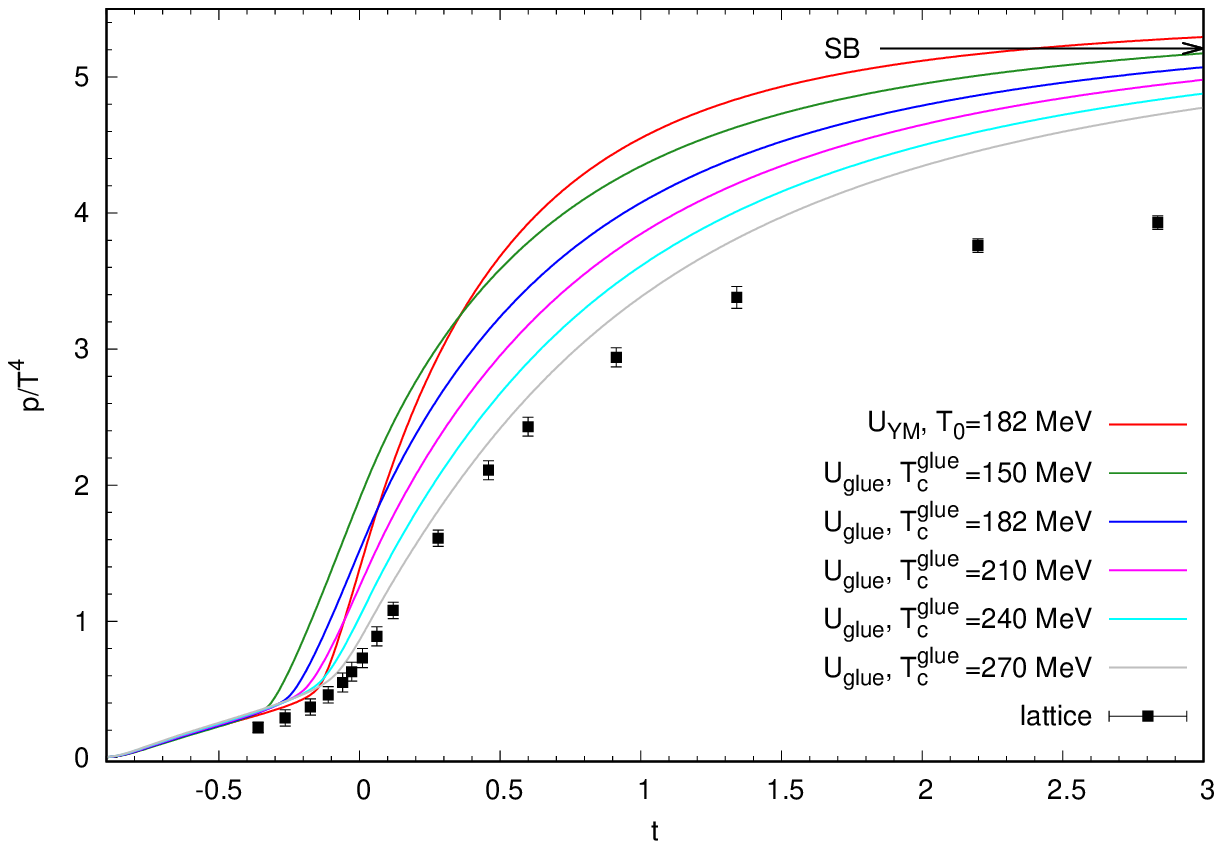}
  \caption{The normalized pressure as a function of the reduced
    temperature $t=T/T_c-1$ and its dependence on the inclusion of various
    mesonic thermal fluctuations in the case when the Polyakov loop
    potential $U_\text{glue}$ is used with $T_c^\text{glue}=270$~MeV (left
    panel) and on the various parametrizations of the Polyakov loop
    potential in the case when the contributions of $\pi, K,$ and $f_0^L$
    are included (right panel).  We compare to the continuum extrapolated
    lattice result of Ref.~\cite{Borsanyi:2010cj}.  The arrow indicates the
    Stefan-Boltzmann limit of the QCD: $p_\text{SB}/T^4=5.209.$}
  \label{Fig:norm_pres_rel_t}
\end{figure*}

In Fig.~\ref{Fig:cond_red_temp} we compare to lattice results the
subtracted chiral condensate \eqref{Eq:sub_chi_cond} obtained by using
the original Polyakov loop potential with the transition temperature
$T_0=182$~MeV and with the improved Polyakov loop potential with
various values of the transition temperature $T_c^\text{glue}.$ In
agreement with the findings of Ref.~\cite{Haas:2013qwp}, one observes
that the chiral transition is slightly smoother when the improved
Polyakov loop potential is used.  The trend of the lattice results is
well reproduced above $T_c$, where the difference between our results
obtained with various Polyakov loop potentials is the smallest.
Comparison with the lattice results at small values seems to favor the
improved Polyakov loop potential with a large value of the
$T_c^\text{glue}$ parameter in the range between 210~MeV and 240~MeV.
In contrast with the nice agreement of the subtracted chiral
condensate with the corresponding lattice result, the thermal
evolution of the Polyakov loop expectation value shown in
Fig.~\ref{Fig:cond_red_temp} is quite far from its lattice
counterpart, as was also the case in Ref.~\cite{Haas:2013qwp}. The
transition shown by the lattice result is much smoother, and although,
as explained in Ref.~\cite{Haas:2013qwp}, the use of the improved
Polyakov loop potential makes the transition smoother compared to the
case when the original logarithmic potential is used, the discrepancy
from the lattice results remains significant. It was argued in
\cite{Andersen:2015sma} that, as the Polyakov loop requires
renormalization, a temperature dependent rescaling has to be applied
to the Polyakov loop expectation value calculated in an effective
model when comparing it to the lattice value.  With this idea the
lattice result of two-color QCD was reproduced in a PNJL
model. Because of the big discrepancy with the lattice data, we could
not apply it in our case, where it seems that the mean field
approximation is rather crude, as far as the expectation value of the
Polyakov loop is concerned.

\subsection{Thermodynamical quantities at $\mu_B=0$}
\label{SubSec:thermodyn}

In this subsection we present the thermodynamical quantities derived
from the pressure and compare them to the corresponding continuum
extrapolated lattice results of Ref.~\cite{Borsanyi:2010cj}. The
pressure is obtained from the grand potential defined in
\eqref{Eq:grand_pot_H} as
\begin{equation}
p (T,\mu_q)=\Omega_\textnormal{H}(T=0,\mu_q) - \Omega_\textnormal{H}(T,\mu_q). 
\end{equation}
Based on the pressure, one can compute thermodynamical observables like the
entropy density $s=\partial p/\partial T,$ the quark number density
$\rho_q=\partial p/\partial \mu_q,$ the quark number susceptibility
$\chi_q=\partial^2 p/\partial \mu^2_q,$ the energy density 
$\epsilon = -p + T s + \mu_q\rho_q,$ as well as the scaled interaction
measure $\Delta=(\epsilon-3 p)/T^4$ and the square of the speed of sound
defined at $\mu_q=0$ as $c^2_s = d p/d\epsilon = s/(T(\partial s/\partial T)).$

So far we have not included any mesonic fluctuations in the grand
potential, and consequently we solved the field equations without
taking them into account. However, the contribution of the pions has
to be included in the pressure, as at small temperature their mass is
the smallest among all constituents of the model. In fact, it is known
from textbooks that for small temperature the scaled pressure behaves
as $p/T^4\sim(m_\pi/T)^{3/2}\exp(-m_\pi/T).$ With the curvature 
mass determined according to Eq.~\eqref{Eq:M2i_ab} from a grand potential not
containing mesonic fluctuations, the additive partial contribution of a meson
$b\in\{\pi, K, f_0^L\}$ to the pressure is taken into account with the formula
\begin{equation}
\Delta p_b(T) = - n_{b}T\int\frac{d^3 p}{(2\pi)^3}\ln(1-e^{-\beta E_b(p)}), 
\end{equation}
where $E_b(p)=\sqrt{p^2 + m_b^2},$ with $m_b$ being the meson mass, and
$n_b$ is the meson multiplicity ($n_\pi=3,$ $n_K=4,$ and $n_{f_0^L}=1$). 
Note that the fermion contribution to the pressure is included using
\eref{Eq:fermi_omega}.

In the left panel of Fig.~\ref{Fig:norm_pres_rel_t} we see that the
constituent quarks and the Polyakov loop potential give the major part of
the contribution to the pressure around and beyond $T_c$ and that at small
temperature the pressure is pion dominated. Any additional mesonic
contribution increases the pressure, and we see that with the inclusion of
$K$ and $f_0^L$, the pressure overshoots the lattice data. The contribution
of the kaons is significant around $T_c$, while that of $f_0^L$ is quite
small in the entire temperature region. This has to do with the
multiplicity of the kaons, as $n_K=4 n_{f_0^L}.$ We included the
contribution of $f_0^L$ in the pressure because in our approximation it
is rather light in the vacuum and its mass decreases with the
temperature roughly up to the pseudocritical temperature $T_c$.

In the right panel of Fig.~\ref{Fig:norm_pres_rel_t} one observes that
with the improved Polyakov loop potential the temperature increase of
the pressure is smoother than with the original Polyakov loop
potential ($U_\text{YM}$), where the Stefan-Boltzmann (SB) limit of
the QCD (ideal gas of massless fermions and gluons) is reached already
for $T\approx 1.5T_c$. One also observes that the overshooting of the
pressure compared with the lattice data, when additional mesonic
contributions are included beyond that of the pions, can be
compensated to some degree by increasing the value of
$T_c^\textnormal{glue}$ in the improved Polyakov loop potential. In
the case of the pressure, we get close to the lattice data by using
the maximal value $T_c^\textnormal{glue}=270$~MeV. This means that one
cannot reproduce equally well all thermodynamical quantities with the
same set of model parameters, as the value for which $\Delta_{l,s}$ is
the closest to the lattice data is---according to
Fig.~\ref{Fig:cond_mass}---in the range $T_c^\textnormal{glue}\in
(210,240)$~MeV. This inconsistency could be related to the
inconsistent treatment of the mesonic contributions which are not
included in the field equations when the model is solved. It is seen
in general that the inclusion of mesonic fluctuations smoothes the
chiral phase transition \cite{Marko_2010,Skokov:2010wb}, and therefore
with their consistent inclusion we would expect a better agreement of
the pressure and the derived thermodynamical quantities with the
lattice results.

\begin{figure}[!t]
\centerline{\includegraphics[width=0.485\textwidth]{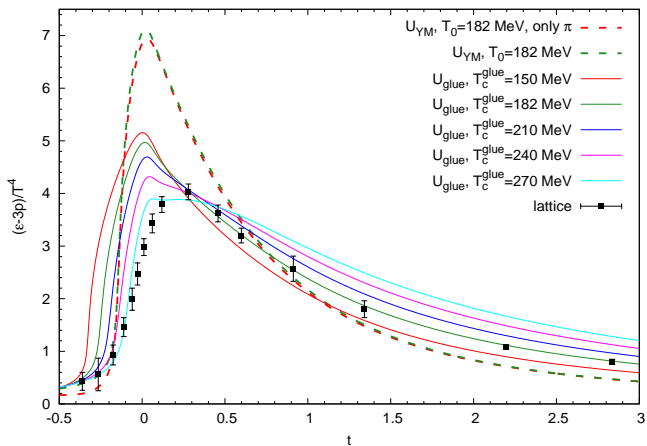}}
\caption{The scaled interaction measure $\Delta = (\epsilon - 3p)/T^4$
  as a function of $t=T/T_c-1$ for various
  parametrizations of the Polyakov loop potential.  Unless indicated
  otherwise, the contribution of $\pi,$ $K,$ and $f_0^L$ is taken into
  account in the grand potential.  The lattice result is from
  Ref.~\cite{Borsanyi:2010cj}.}
\label{Fig:interaction_measure}
\end{figure}

\begin{figure}
\centerline{\includegraphics[width=0.485\textwidth]{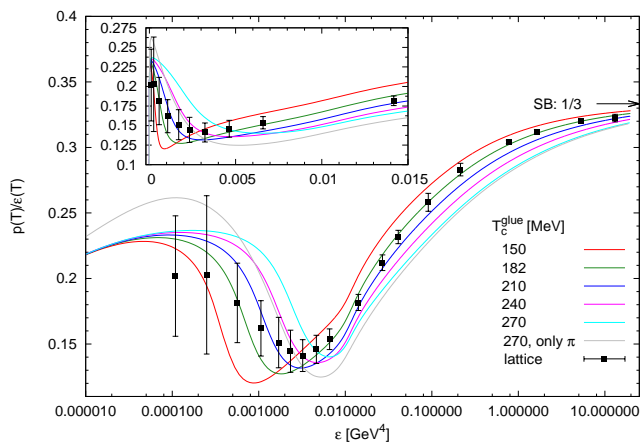}}
\caption{The ratio $p/\epsilon$ as a function of the energy density
  obtained using the improved Polyakov loop potential at different
  values of $T^\text{glue}_c.$ Unless indicated otherwise, the
  contribution of $\pi,$ $K,$ and $f_0^L$ is taken into account in the
  grand potential. Note the logarithmic scale of the abscissa in the
  main plot. The inset zooms into the region of small $\epsilon.$ }
\label{Fig:EOS_thermal_contr}
\end{figure}

\begin{figure}[!t]
\centerline{\includegraphics[width=0.485\textwidth]{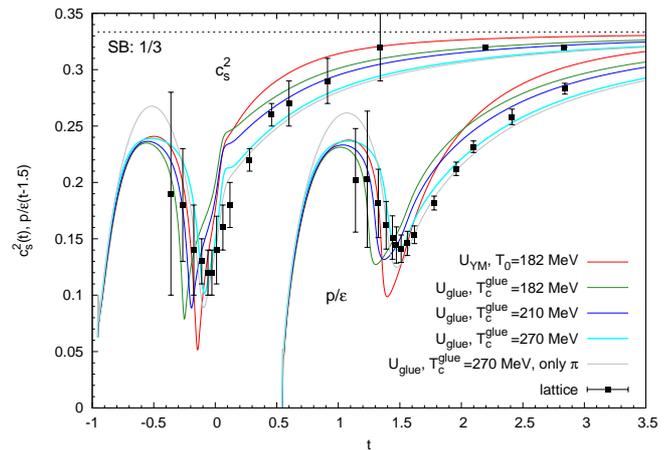}}
\caption{The square of the speed of sound $c_s^2$ and the ratio
  $p/\epsilon$ as functions of the reduced temperature $t=T/T_c-1$
  for various parametrizations of the Polyakov loop potential. Note
  that $p/\epsilon$ is shifted to the right by 1.5. Unless indicated
  otherwise, the contribution of $\pi,$ $K,$ and $f_0^L$ is taken into
  account in the grand potential. }
\label{Fig:sound_T_gl}
\end{figure}

Some thermodynamical quantities derived from the pressure, like the
scale interaction measure $\Delta,$ the square of the speed of sound
$c_s^2$, and the equation of state parameter $p/\epsilon$ (pressure
over energy density), are presented in
Figs.~\ref{Fig:interaction_measure}, \ref{Fig:EOS_thermal_contr}, and
\ref{Fig:sound_T_gl}.  With the original Polyakov loop potential the
maximum of the scaled interaction measure $\Delta(t)$ in
Fig.~\ref{Fig:interaction_measure} and the minimum of the square of
the speed of sound $c_s^2(t)$ in Fig.~\ref{Fig:sound_T_gl}
($t=T/T_c-1$ is the reduced temperature) turn out to be too high and
too low, respectively.  With the improved Polyakov loop potential the
trend of the corresponding continuum extrapolated lattice results are
fairly well reproduced by our results.  As far as the mesonic sector
is concerned, the presented quantities are basically pion dominated;
however, the lattice results are better reproduced if the contributions
of kaons and $f_0^L$ are taken into account. One observes in
Figs.~\ref{Fig:EOS_thermal_contr} and \ref{Fig:sound_T_gl} that at
high temperature both $c_s^2$ and $p/\epsilon$ approach $1/3,$ which
is the value obtained in the Stefan-Boltzmann limit of the QCD.

We also studied the effect of the transition temperature
$T_c^\text{glue}$ appearing in the improved Polyakov loop
potential. As it is visible from the figures, a single value of
$T_c^\text{glue}$ cannot reproduce equally well the values of all the
quantities determined on the lattice. In the case of the interaction
measure a large value of $T_c^\text{glue}$ is favored around $T_c$ and
a smaller one at large temperatures. The minimum of $c_s^2(t)$ and
$p(t)/\epsilon(t)$ is well described with
$T_c^\text{glue}\simeq270$~MeV, while the minimum of $p/\epsilon$
plotted as a function of $\epsilon$ is fairly well reproduced with a
different value, $T_c^\text{glue}\simeq210$~MeV.

\subsection{$\mu_B-T$ phase diagram and the critical endpoint}
\label{SubSec:phase_diag}

We turn now to the study of the chiral phase transition at finite
baryon chemical potential $\mu_B.$ As $\mu_B$ is increased from zero,
the chiral transition as a function of the temperature becomes more
and more rapid, although its crossover nature is preserved for quite
large values of $\mu_B$. The pseudocritical temperature decreases with
increasing $\mu_B$ and one can determine at $\mu_B=0$ the curvature
$\kappa$ of the chiral crossover transition curve in the $T-\mu_B$
plane through the following standard fit
\begin{eqnarray}
\frac{T_c(\mu_B)}{T_c({\mu_B=0})} = 1 - \kappa \left(\frac{\mu_B}{T_c(\mu_B)}\right)^2.
\end{eqnarray}
We obtain $\kappa=0.0193,$ which is very close to the continuum extrapolated
lattice result $\kappa=0.020(4)$ reported in \cite{Cea:2015cya} for the
case $\mu_u=\mu_d=\mu_s.$\footnote{We thank M. D'Elia for indicating the
appropriate reference to compare our result with.} We mention that when 
$\mu_u=\mu_d$ and $\mu_s=0$, the lattice results are significantly
lower: $\kappa=0.0135(20)$ in \cite{Bonati:2015bha} and $\kappa=0.0149(21)$ 
in \cite{Bellwied:2015rza}.

\begin{figure}[!t]
\centerline{\includegraphics[width=0.485\textwidth]{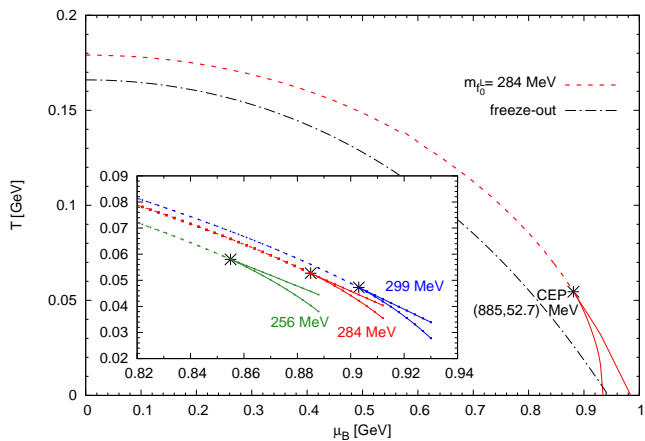}}
\caption{The phase diagram obtained by using the improved Polyakov
  loop potential $U_\text{glue}$ with $T^\text{glue}_c=210$~MeV. The inset
  shows the dependence of the CEP's location on $m_{f_0^L}.$ The dashed curve
  denotes a crossover-type transition, the solid curves represent the two
  spinodals limiting the metastable region associated with a first order phase
  transition, while the dash-dotted curve is the chemical freeze-out curve
  of Ref.~\cite{Cleymans:2006qe} (see text for details).}
\label{Fig:phase_bound}
\end{figure}

In the case of our best set of parameters determined in
Sec.~\ref{Sec:parametrization}, the crossover transition eventually
turns with increasing $\mu_B$ into a first order one, by passing
through the CEP of the phase boundary, where the transition is second
order.  This is presented in Fig.~\ref{Fig:phase_bound}, where we show
the phase diagram obtained with the improved Polyakov loop potential
\eqref{Eq:U_glue} by using $T_c^\text{glue}=210$~MeV.  The crossover
transition curve can be parametrized as $T_c(\mu_B) =
T_c(\mu_B=0)-0.101\mu_B^2-0.073\mu_B^4$ with
$T_c(\mu_B=0)=0.179$~GeV. Since it was argued in
\cite{BraunMunzinger:2003zz} that the chemical freeze-out temperature
is close to the critical temperature, it is interesting to compare the
above transition curve with the chemical freeze-out curve deduced from
particle multiplicities in heavy ion collisions, to which the
parametrization $T=0.166-0.139\mu_B^2-0.053 \mu_B^4$ was given in
\cite{Cleymans:2006qe}, with $T$ and $\mu_B$ measured in GeV.  One can
see in Fig.~\ref{Fig:phase_bound} that our $T_c(\mu_B)$ phase
transition curve lies farther from the origin of the $T-\mu_B$ plane
than the freeze-out curve.

With the best set of parameters given in Table~\ref{Tab:param}, the
location of the CEP in our model is given by
$(\mu_B^\text{CEP},T_c^\text{CEP})=(885,52.7)$~MeV when the improved
Polyakov loop potential is used with $T^\text{glue}_c=210$~MeV. The
phase diagram is similar to that obtained in \cite{Stiele:2016cfs},
with comparable values of the CEP's coordinates, and we refer to that
paper concerning the influence of the improvement in the Polyakov loop
potential on the location of the CEP in the PL$\sigma$M. The large
value of $\mu_B^\text{CEP}$ we obtained is typical of a linear sigma
model without (axial)vector mesons in the case when the vacuum
fluctuation of the fermions is included. See, e.g.,
\cite{Chatterjee:2011jd} where the value $\mu_B^\text{CEP}=849$~MeV
was reported, with a somewhat larger value of temperature,
$T_c^\text{CEP}=81$~MeV, than is in our case. Without the inclusion of
the fermionic vacuum fluctuations, as is the case of
Refs.~\cite{Schaefer:2008hk,Mao:2009aq}, the value of
$\mu_B^{\text{CEP}}$ is smaller and $T_c^{\text{CEP}}$ higher,
compared to the case when they are properly taken into account. In
Fig.~7 of \cite{Chatterjee:2011jd} one can see that the inclusion of
the fermionic vacuum fluctuations shifts the CEP found for
$m_\sigma=400$~MeV at
$(\mu_B^\text{CEP},T_c^\text{CEP})=(240,177.5)$~MeV to the location
quoted above, that is $\mu_B^\text{CEP}/T_c^\text{CEP}$ increases from
$1.35$ to $10.5.$

The continuum extrapolated lattice results of \cite{Bellwied:2015rza},
obtained using analytical continuation from imaginary chemical
potential, show no evidence of CEP up to $\mu_B\approx
350$~MeV. Although there exist lattice estimates on the location of
the CEP, these are obtained at fixed lattice spacing and temporal
extent $N_t=4$ and at different numbers of flavors ($N_f=2$ and
$N_f=2+1$), value of the pion mass, and lattice volume (see Table I of
\cite{Gavai:2004sd}). Opposed to these is the lattice result
\cite{deForcrand:2006pv} obtained at $N_t=4$, in which the shrinking
of the first order chiral transition region of the $m_{u,d}-m_s$ plane
was observed as $\mu_B$ is increased from zero. The result of
Ref.~\cite{deForcrand:2006pv} would suggest the absence of CEP, unless
the $\mu_B^\text{crit}(m_{u,d},m_s)$ surface of the second order phase
transition points behaves nonmonotonously with increasing $\mu_B,$
similar to the situation observed, e.g., in \cite{Chen:2009gv} in the
Nambu-Jona-Lasinio model, using a $\mu_B$-dependent 't Hooft
coupling, or in \cite{Jakovac:2010uy}, in the linear sigma model.

Instead of comparing the location of the CEP found in our model to
lattice results obtained at fixed lattice spacing, we compare it with
values coming from the solutions of truncated Dyson-Schwinger
equations in Landau gauge QCD obtained with $N_f=2$ \cite{Qin:2010nq,
  Shi:2014zpa, Gutierrez:2013sta} and with $N_f=2+1$
\cite{Fischer:2014ata,Eichmann:2015kfa}, and also with an estimate
obtained by analyzing experimental data in heavy-ion collisions
\cite{Lacey:2014wqa}. Simple parametrizations of the gluon propagator
gives $\mu_B^\text{CEP}/T_c^\text{CEP}\simeq 3.3$ in
\cite{Qin:2010nq}, which does not seem to depend on the dressing of
the quark-gluon vertex, and $\mu_B^\text{CEP}/T_c^\text{CEP}\simeq
3.4$ in \cite{Shi:2014zpa}. On the other hand, when a temperature
dependent parametrization of the gluon propagator is used in
\cite{Gutierrez:2013sta}, based on which the $T$ dependence of the
quark-antiquark condensate is reproduced at $\mu_B=0,$
$\mu_B^\text{CEP}/T_c^\text{CEP}\simeq 6.8$ is obtained, which is a
factor of $2.5$ smaller than our value and a factor of 2 larger than
the values in \cite{Qin:2010nq, Shi:2014zpa}. Compared to these
values, $\mu_B^\text{CEP}/T_c^\text{CEP}\simeq 4.4$ was found in
\cite{Fischer:2014ata} (our value of $\mu_B^\text{CEP}$ is 1.75 times
larger and our value of $T_c^\text{CEP}$ is $2$ times smaller than
there), which increases slightly to $4.7$ \cite{Eichmann:2015kfa} with
the inclusion of terms in the quark-gluon interaction which are
parametrized with baryonic degrees of freedom. What is common in the
approach based on the Dyson-Schwinger equations and also in the method
of \cite{Ayala:2011vs}\footnote{We thank the referee for bringing this
  reference and also Ref.~\cite{Gutierrez:2013sta} to our
  attention. Note that in \cite{Ayala:2011vs} no CEP was found for
  $\mu_B\le 0.3$~GeV, that is, in the range of $\mu_B$ where the
  approximation used is valid.} using finite energy sum rules is that
they rely on the self-consistent propagator equation for the
quarks. It remains to be seen how self-energy correction in the
fermion propagator will affect in our model the value of
$\mu_B^\text{CEP}/T_c^\text{CEP}.$

\begin{figure*}
  \includegraphics[width=0.485\textwidth]{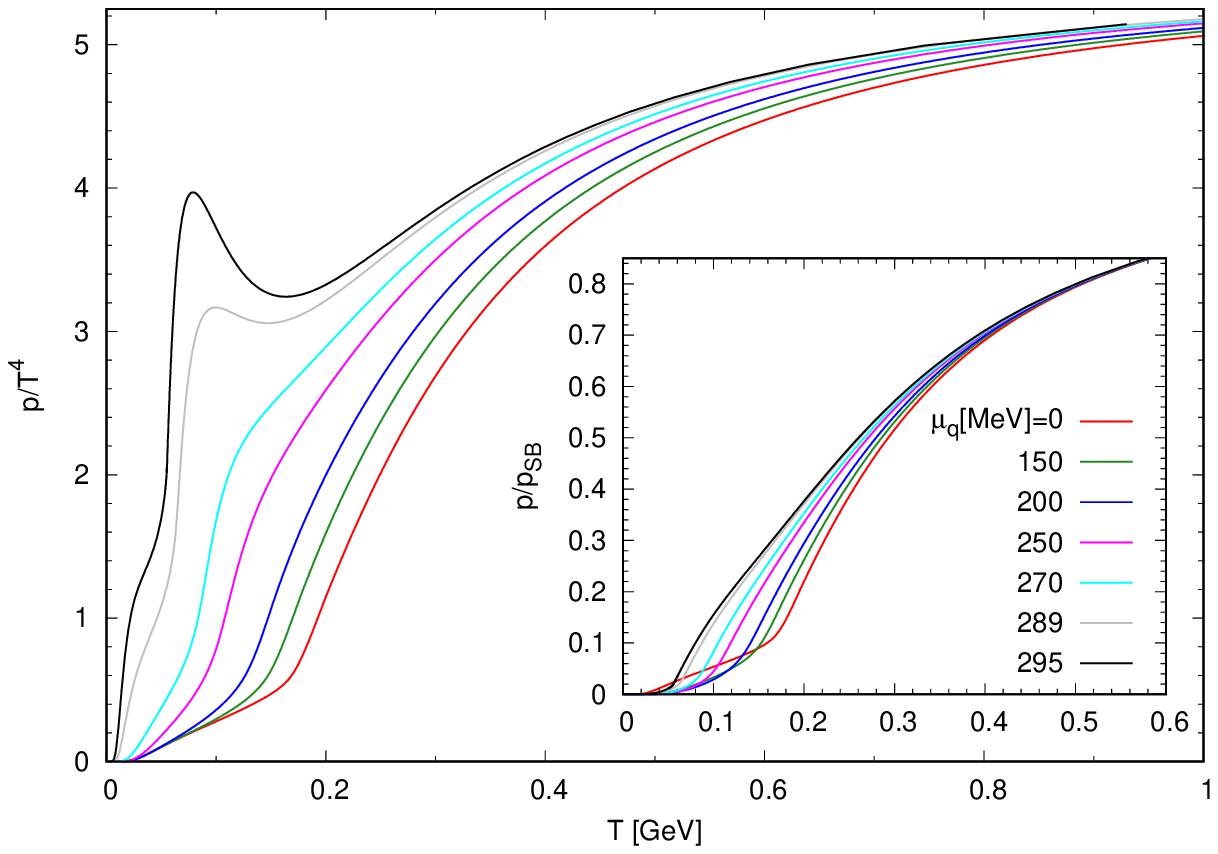}\quad
  \includegraphics[width=0.485\textwidth]{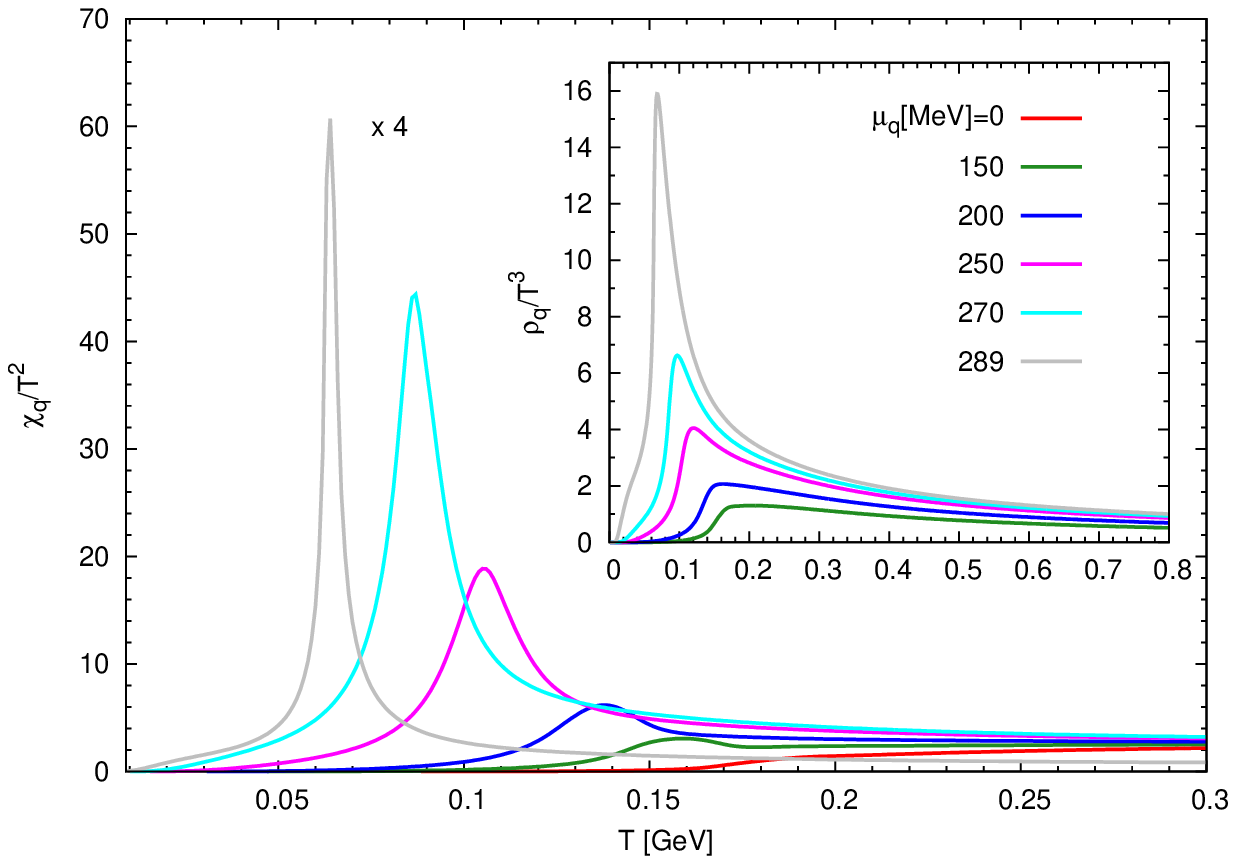}
  \vspace*{0.25cm}
  
  \includegraphics[width=0.485\textwidth]{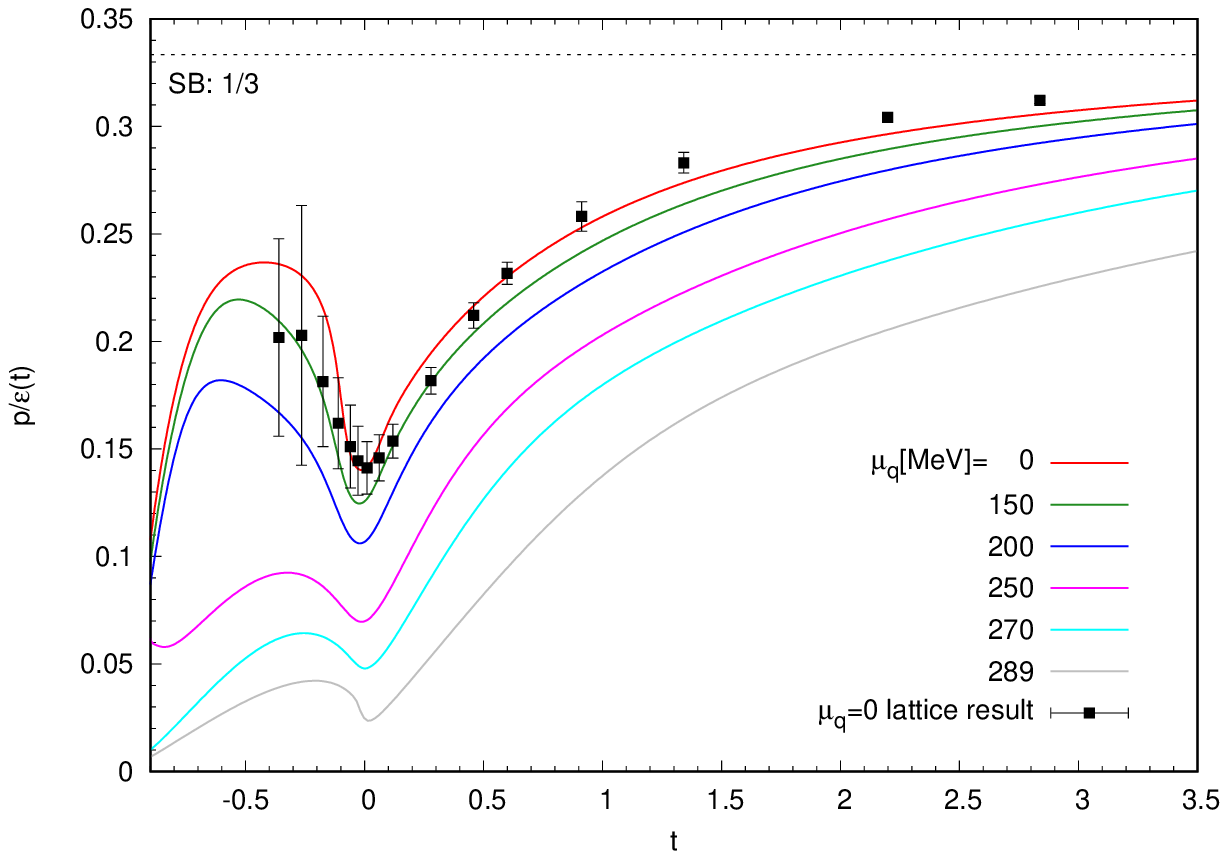}\quad 
  \includegraphics[width=0.485\textwidth]{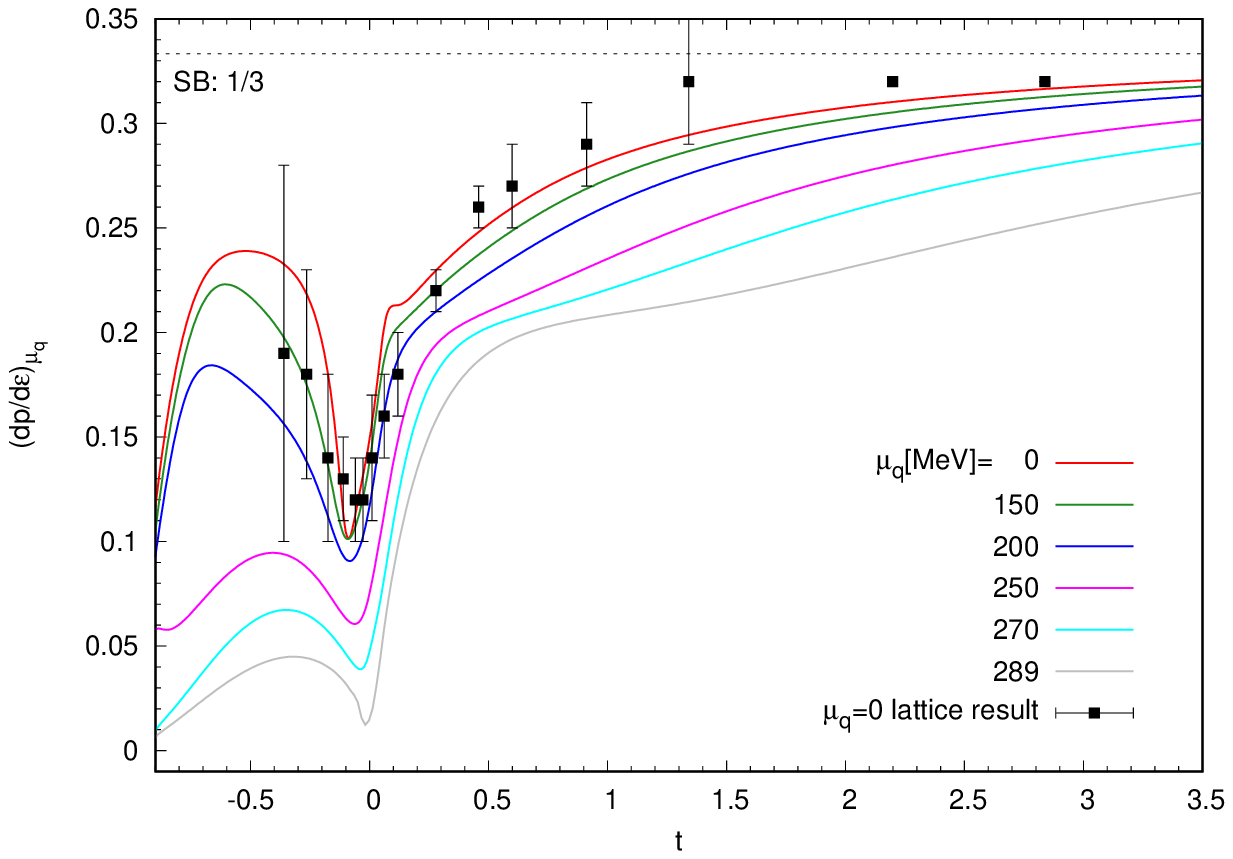}
  \caption{The temperature dependence of thermodynamical observables
    at various values of quark chemical potential $\mu_q$ obtained by
    using the improved Polyakov loop potential $U_\text{glue}$ with
    $T^\text{glue}_c=270$~MeV and by taking into account the
    contribution of $\pi,$ $K,$ and $f_0^L$ in the pressure. In the
    top row we show the scaled pressure $p/T^4$ (left main plot) and
    $p/p_{SB}$ (inset), scaled quark number susceptibility
    $\chi_q/T^2$ (right main plot), and the scaled quark number density
    $\rho_q/T^3$ (inset). In the bottom row we present the ratio
    $p/\epsilon$ and the quantity $(d p/d\epsilon)_{\mu_q},$ defined
    in the main text, as functions of the reduced temperature
    $t=T/T_c-1.$ }
\label{Fig:therm_obs_mu_B}
\end{figure*}

Recently the nonmonotonic pattern of some experimental observable
obtained at various centralities as a function of the collision energy
was attributed in \cite{Lacey:2014wqa} to finite-size scaling effects
occurring near a second order phase transition. The determined
critical exponents governing the growth of the correlation length and
susceptibility suggests the existence of a CEP that belongs to the
universality class of a three-dimensional Ising model and has a small value of baryon
chemical potential, $\mu_B^\text{CEP}\approx 95$~MeV, and a high value
of critical temperature $T_c^\text{CEP} \approx 165$~MeV. 

In the inset of Fig.~\ref{Fig:phase_bound} we show the variation of
the CEP's location with the value of the $\sigma\equiv f_0^L$ mass.
Increasing values of the mass push the position of the CEP to higher
values of $\mu_B$ and lower values of the temperature, as was observed
previously in the literature in cases when only the scalar and
pseudoscalar mesons were incorporated in the model, both without or
with the inclusion of the fermionic vacuum fluctuations; see
\cite{Schaefer:2008hk} and \cite{Chatterjee:2011jd}, respectively. In
our case, it turned out that there is no CEP when the value of
$m_{f_0^L}$ is pushed beyond $\approx 340$~MeV.
For changing the $f_0^L$ mass artificially we increased the weight of
the $f_0^L$ mass in the $\chi^2$ fit from $1$ to $20$, which
forced the fit to reach the desired mass value. We set multiple
values in the $220-500$~MeV mass range. Even if we increased the
mass weight to $20$ the resulting $f_0^L$ mass could differ from its
prescribed value significantly. Finally, we ended up with two
additional distinct mass values ($256$~MeV and $299$~MeV) for which the
CEP exist, as shown in Fig.~\ref{Fig:phase_bound}.

\subsection{Thermodynamical quantities at $\mu_B\ne0$}

In the previous subsection we have located the CEP in the $T-\mu_B$
phase diagram by monitoring the temperature evolution of the
nonstrange condensate at increasing values of the baryon chemical
potential $\mu_B.$ Now we present in Fig.~\ref{Fig:therm_obs_mu_B} the
temperature evolution of various thermodynamical observables at
increasing values of $\mu_q=\mu_B/3,$ as the CEP is approached from
the region of the phase diagram where the chiral transition is an
analytic crossover. Some of these observables have a peculiar behavior
in the vicinity of a second order phase transition, some others
increase and diverge at the CEP and therefore in principle they can be
used in an experimental setting to signal its presence.

We see in Fig.~\ref{Fig:therm_obs_mu_B} that the presence of the CEP is
signaled by the nonmonotonic temperature dependence of the scaled pressure
$p/T^4.$ If the pressure is scaled with the pressure of the QCD in the SB
limit, namely, by
\begin{eqnarray}
p_{SB}(T,\mu_q)&=&(N_c^2-1)\frac{\pi^2}{45}T^4 \nonumber\\
&+& N_c N_f
\left[\frac{7\pi^2}{180}T^4+\frac{T^2\mu_q^2}{6}+\frac{\mu_q^4}{12\pi^2}\right]\,,\ \ 
\end{eqnarray}
where $N_c$ and $N_f$ are, respectively, the number of colors and
flavors, then the presence of the CEP is hardly visible.  The effect
of the CEP appears magnified in the scaled quark number susceptibility
$\chi_q/T^2$ and the scaled quark number density $\rho_q/T^3.$ Note
that by increasing $\mu_q$ from 270~MeV to 289~MeV, which is very
close to the coordinate of the CEP, the value of the scaled quark
number susceptibility is increased by a factor of 4 (for the sake of the
presentation we divided by this factor the value of $\chi_q/T^2$
obtained at $\mu_q=289$~MeV).

In the bottom row of Fig.~\ref{Fig:therm_obs_mu_B} we show at several
fixed values of $\mu_q$ the temperature variation of $p/\epsilon$ and
$(d p/d\epsilon)_{\mu_q} = s/(T(\partial s/\partial T) +\mu_q
(\partial\rho_q/\partial T)).$ This quantity, derived at constant
$\mu_q,$ connects at $\mu_q=0$ with the square of the speed of sound
$c_s^2.$ One sees that both $p/\epsilon$ and $(d p/d\epsilon)_{\mu_q}$
decrease with increasing values of $\mu_q,$ and that the minimum of
the latter quantity approaches zero at CEP and shows a very steep
rise, as the temperature increases above the critical value.

\section{Conclusions}
\label{Sec:conclusion}

We have studied at finite temperature and baryonic densities the
thermodynamical properties of the Polyakov loop extended quark meson
model containing also vector and axial vector mesons. These latter
ingredients manifest themselves in a nontrivial way in the vacuum
parametrization of the model through their tree-level masses and decay
widths.  The $\chi^2$-minimization procedure applied earlier in
\cite{elsm_2013} was modified by including the effect of the fermion
vacuum fluctuations in the scalar and pseudoscalar meson curvature
masses and also by considering as an input the well-established value
of the chiral transition temperature of the QCD at vanishing density.

With our parametrization procedure we have investigated which scalar
particles with mass below 2~GeV can be assigned to the scalar states
of the model, under the assumption that these are $q\bar q$ states. It
turned out that the smallest value of $\chi^2$ is reached when the
states of the model correspond to $a_0(980),$ $K_0^\star(800),$
$f_0(500),$ and $f_0(980)$ particles. For this particular particle
assignment we have studied the thermodynamics of the model, by using
an improved Polyakov loop potential, recently proposed in the
literature, and found that a CEP of the crossover transition line
exists in the $T-\mu_B$ phase diagram at rather large values of
$\mu_B.$ We have computed various thermodynamical observables and
compared them with continuum extrapolated lattice results.  Based on
the fairly good agreement with the lattice data observed at vanishing
density, it would be interesting to use in astrophysical applications
the finite density equation of state of our model.

The inclusion of the pseudocritical temperature in the parametrization
procedure proved crucial, as it drastically reduced the number of acceptable
solutions of the $\chi^2$-minimization procedure.  It turned out that in
order for the model to provide a meaningful thermodynamics, $f_0(500)$ has
to be part of the scalar multiplet, in contrast with the parametrization
based exclusively on mesonic vacuum quantities \cite{elsm_2013}, where, in
the absence of fermionic vacuum fluctuations, $f_0(1370)$ and $f_0(1710)$
were found to belong to the scalar nonet states.  This contradiction is most
probably a consequence of the fact that in both cases the parametrization of
the model was done as if the scalar states were all $\bar q q$ excitations,
which is more likely not the case in nature.  Therefore, it would be
interesting to consider in the future the mixing of the $\bar q q$ states
with tetraquark states and redo the parametrization of the model and the
thermodynamical investigation presented here.  We mention that additional
details on the properties of some scalar particles recently given in
\cite{Pelaez:2015qba, Wolkanowski:2015lsa, Wolkanowski:2015jtc,
Fariborz:2015era} can also be considered in a future work.  Moreover, beside
the tetraquarks, the glueball admixture presumably existing in some
components of the isoscalar sector \cite{Janowski:2011gt, Janowski:2014ppa}
also has to be taken into account because, as a result of the mixing of the
isoscalar states, this admixture can influence the results presented here.

\section*{Acknowledgments}

The authors were supported by the Hungarian Research Fund (OTKA) under
Contract No.  K109462 and by the HIC for FAIR Guest Funds of the Goethe
University Frankfurt.  Zs.~Sz.  would like to thank Anja Habersetzer for
many valuable discussions which gave him insight into various technical and
physical aspects of the EL$\sigma$M.  P.~K.  and Zs.~Sz.  would like to
thank P{\'e}ter V\'an and Antal Jakov\'ac, respectively, for helpful
discussions concerning thermodynamical observables.

\appendix

\section{Scalar decay widths}
\label{App:decays}

In this appendix the expressions of the scalar decays are listed,
which are used in the parametrization and have been changed compared
to \cite{elsm_2013} due to the change of the anomaly term ($\propto
c_1$) in \eref{Eq:Lagr}. The affected decays are $\Gamma_{a_0}$ (this
is the sum of the following three decay widths:
$\Gamma_{a_0\to\eta\pi}$, $\Gamma_{a_0\to\eta^{\prime}\pi}$, and
$\Gamma_{a_0\to K K}$), $\Gamma_{K_0^{\star}\to K \pi}$,
$\Gamma_{f_0^{L/H}\to\pi\pi}$, and $\Gamma_{f_0^{L/H}\to K K}$.  

The tree-level $a_{0}\rightarrow\eta\pi$ as well as $a_{0}\rightarrow
\eta^{\prime} \pi$ decay widths read
\begin{align}
\Gamma_{a_{0}\rightarrow\eta\pi}  &  =\frac{1}{8m_{a_{0}}\pi}\left[
\frac{(m_{a_{0}}^{2}-m_{\eta}^{2}-m_{\pi}^{2})^{2}-4m_{\eta}^{2}m_{\pi}^{2}
}{4m_{a_{0}}^{4}}\right]  ^{1/2}\nonumber \\
&\quad \times |\mathcal{M}_{a_{0}\rightarrow\eta\pi}|^{2},\label{B1}\\
\Gamma_{a_{0}\rightarrow\eta^{\prime}\pi}  &  =\frac{1}{8m_{a_{0}}\pi}\left[
\frac{(m_{a_{0}}^{2}-m_{\eta^{\prime}}^{2}-m_{\pi}^{2})^{2}-4m_{\eta^{\prime}
}^{2}m_{\pi}^{2}}{4m_{a_{0}}^{4}}\right]  ^{1/2}\nonumber \\
&\quad \times |\mathcal{M}_{a_{0}\rightarrow\eta^{\prime}\pi}|^{2}, \label{B2}
\end{align}
with the following transition matrix elements,
\begin{align}
\mathcal{M}_{a_{0}\rightarrow\eta\pi}  &  =\cos\theta_{\pi}\mathcal{M}_{a_{0}\rightarrow\eta_{N}\pi}(m_{\eta}) \nom\\
&\quad +\sin\theta_{\pi}\mathcal{M}_{a_{0}\rightarrow\eta_{S}\pi}(m_{\eta})\;,\\
\mathcal{M}_{a_{0}\rightarrow\eta^{\prime}\pi}  &  =\cos\theta_{\pi
}\mathcal{M}_{a_{0}\rightarrow\eta_{S}\pi}(m_{\eta^{\prime}})\nom \\
&\quad -\sin\theta_{\pi}\mathcal{M}_{a_{0}\rightarrow\eta_{N}\pi}(m_{\eta^{\prime}})\;,
\end{align}
where
\begin{align}
\mathcal{M}_{a_{0}\rightarrow\eta_{N}\pi}(m)  &  =A_{a_{0}\eta_{N}\pi
}-B_{a_{0}\eta_{N}\pi}\frac{m_{a_{0}}^{2}-m^{2}-m_{\pi}^{2}}{2}\nom \\
&\quad +C_{a_{0}\eta_{N}\pi}m_{a_{0}}^{2},\\
\mathcal{M}_{a_{0}\rightarrow\eta_{S}\pi}(m)  &  =A_{a_{0}\eta_{S}\pi},
\end{align}
and
\begin{align}
A_{a_{0}\eta_{N}\pi}  &  =-Z_{\pi}^{2}\lambda_{2}\phi
_{N},\\
B_{a_{0}\eta_{N}\pi}  &  =-2\frac{g_{1}^{2}\phi_{N}}{m_{a_{1}}^{2}}\left[
1-\frac{1}{2}\frac{Z_{\pi}^{2}\phi_{N}^{2}}{m_{a_{1}}^{2}}(h_{2}%
-h_{3})\right]  ,\\
C_{a_{0}\eta_{N}\pi}  &  =g_{1}Z_{\pi}^{2}w_{a_{1}},\\
A_{a_{0}\eta_{S}\pi}  &  =\frac{1}{2}c_{1}Z_{\pi}Z_{\eta_{S}}\phi_{N}^{2}%
\phi_{S}.
\end{align}
 
The $a_{0}\rightarrow KK$ decay width is found to be
\begin{align}
\Gamma_{a_{0}\rightarrow KK}&=\frac{1}{8m_{a_{0}}\pi}\sqrt{1-\left(
\frac{2m_{K}}{m_{a_{0}}}\right)  ^{2}}\bigg\vert A_{a_{0}KK} \nom \\ 
&\quad  -\frac{1}{2}B_{a_{0}KK}(m_{a_{0}}^{2}-2m_{K}^{2})+C_{a_{0}KK}m_{a_{0}}^{2}
\bigg\vert^{2}\;, \label{B3}
\end{align}
where
\begin{align}
A_{a_{0}KK}  &  =Z_{K}^{2}\left(\lambda_{2}\left( \phi_{N}-\frac{\phi_{S}}{\sqrt
{2}}\right) + \frac{1}{4}c_1\right)  ,\\
B_{a_{0}KK}  &  =Z_{K}^{2}w_{K_{1}}\bigg\{  g_{1}-\frac{1}{2}w_{K_{1}}(
(g_{1}^{2}+h_{2})\phi_{N} \nom \\
&\quad  +\sqrt{2}(g_{1}^{2}-h_{3})\phi_{S})  \bigg\},\\
C_{a_{0}KK}  &  =-\frac{g_{1}}{2}Z_{K}^{2}w_{K_{1}}.
\end{align}
It is worth noting that in the expressions above only the forms of
$A_{a_{0}\eta_{N}\pi}$ and $A_{a_{0}KK}$ have changed.

Now turning to the scalar kaon, the decay width reads
\begin{eqnarray}
\Gamma_{K_0^{\star}\rightarrow K\pi}&=&\frac{3}{8\pi m_{K_{0}^{\star}}}\left[  \frac
{(m_{K_{0}^{\star}}^{2}-m_{\pi}^{2}-m_{K}^{2})^{2}-4m_{\pi}^{2}m_{K}^{2}}{4m_{K_{0}^{\star}}^{4}}
\right]^{1/2}  \nom \\
&\times& \left[\frac{}{}A_{K_{0}^{\star}K\pi}+\frac{1}{2}(C_{K_{0}^{\star}K\pi}+D_{K_{0}^{\star}K\pi}-B_{K_{0}^{\star}K\pi}) \right.\nom\\
&&\  \times\big(m_{K_{0}^{\star}}^{2}-m_{K}^{2}-m_{\pi}^{2}\big)+C_{K_{0}^{\star}K\pi}m_{K}^{2}\nom \\
&&\  + D_{K_{0}^{\star}K\pi}m_{\pi}^{2}\bigg]\;,
\end{eqnarray}
with
\begin{align}
A_{K_{0}^{\star}K\pi}  &
=Z_{\pi}Z_{K}Z_{K_{0}^{\star}}\left(\lambda_{2}\frac{\phi_{S}}{\sqrt{2}} +
  \frac{c_1}{2} \right),\\
B_{K_{0}^{\star}K\pi}  &  =\frac{Z_{\pi}Z_{K}Z_{K_{0}^{\star}}}{4}w_{a_{1}}w_{K_{1}}\left[
2g_{1}\frac{w_{a_{1}}+w_{K_{1}}}{w_{a_{1}}w_{K_{1}}}\right.\nom \\
&\quad \left.+(2h_{3}-h_{2}-3g_{1}^{2})\phi_{N}-\sqrt{2}(g_{1}^{2}+h_{2})\phi_{S})\right],\\
C_{K_{0}^{\star}K\pi}  &
=\frac{Z_{\pi}Z_{K}Z_{K_{0}^{\star}}}{2}[-g_{1}(iw_{K^{\star}} + w_{K_{1}})\nom \\
&\quad +\sqrt{2}iw_{K_0^{\star}}w_{K_{1}}(g_{1}^{2}-h_{3})\phi_{S}],\\
D_{K_{0}^{\star}K\pi}  &  =\frac{Z_{\pi}Z_{K}Z_{K_{0}}}{4}[2g_{1}(iw_{K_0^{\star}
}-w_{a_{1}})+iw_{K_0^{\star}}w_{a_{1}} \nom \\
&\quad \times ((2h_{3}-h_{2}-3g_{1}^{2})\phi_{N} +\sqrt{2}(g_{1}^{2}+h_{2})\phi_{S})],
\end{align}
where only $A_{K_{0}^{\star}K\pi}$ has changed. 

The decay widths of the $f_{0}^{{L/H}}$ in the $\pi\pi$ channel are
\begin{align}
\Gamma_{f_{0}^{{L}}\rightarrow\pi\pi}  &  =\frac{3}{32\pi m_{f_{0}%
^{\text{L}}}}\sqrt{1-\left(  \frac{2m_{\pi}}{m_{f_{0}^{\text{L}}}}\right)
^{2}}\left\vert \mathcal{M}_{f_{0}^{{L}}\rightarrow\pi\pi}\right\vert
^{2}\;,\\
\Gamma_{f_{0}^{{H}}\rightarrow\pi\pi}  &  =\frac{3}{32\pi m_{f_{0}%
^{H}}}\sqrt{1-\left(  \frac{2m_{\pi}}{m_{f_{0}^{H}}}\right)
^{2}}\left\vert \mathcal{M}_{f_{0}^{{H}}\rightarrow\pi\pi}\right\vert
^{2}\;,
\end{align}
where the matrix elements are
\begin{align}
\mathcal{M}_{f_{0}^{{L}}\rightarrow\pi\pi}  &  =-\sin\theta_{\sigma
}\mathcal{M}_{f_{0}\pi}^{{H}}(m_{f_{0}^{{L}}})+\cos\theta_{\sigma
}\mathcal{M}_{f_{0}\pi}^{{L}}(m_{f_{0}^{{L}}}),\\
\mathcal{M}_{f_{0}^{{H}}\rightarrow\pi\pi}  &  =\cos\theta_{\sigma
}\mathcal{M}_{f_{0}\pi}^{{H}}(m_{f_{0}^{{H}}})+\sin\theta_{\sigma
}\mathcal{M}_{f_{0}\pi}^{{L}}(m_{f_{0}^{{H}}}),\\
\mathcal{M}_{f_{0}\pi}^{{L}}(m)  &  =2Z_{\pi}^{2}\phi_{N}\left\{
\frac{g_{1}^{2}}{2}\frac{m^{2}}{m_{a_{1}}^{2}}\left[  1+\left(  1-\frac
{2m_{\pi}^{2}}{m^{2}}\right)\right.\right.\nom\\
& \left.\left.\quad\times \frac{m_{1}^{2}+h_{1}\phi_{S}^{2}/2+2\delta_{N}%
}{m_{a_{1}}^{2}}\right]  -\left(  \lambda_{1}+\frac{\lambda_{2}}{2}\right)
\right\}  ,\label{aa}\\
\mathcal{M}_{f_{0}\pi}^{{H}}(m)  &  =2Z_{\pi}^{2}\phi_{S}\left\{
-\frac{g_{1}^{2}}{4}\frac{m^{2}}{m_{a_{1}}^{2}}\left(  1-\frac{2m_{\pi}^{2}%
}{m^{2}}\right)  \frac{h_{1}\phi_{N}^{2}}{m_{a_{1}}^{2}}\right.\nom\\
& \left. \quad-\lambda_{1} + \frac{c_1}{2\sqrt{2}\phi_S}\right\}
. \label{bb}
\end{align}
In the $KK$ channel the decay widths read
\begin{align}
\Gamma_{f_{0}^{H}\rightarrow KK}  &  =\frac{1}{8\pi m_{f_{0}^{H%
}}}\sqrt{1-\left(  \frac{2m_{K}}{m_{f_{0}^{H}}}\right)  ^{2}}\left\vert
\mathcal{M}_{f_{0}^{H}\rightarrow KK}\right\vert ^{2},\\
\Gamma_{f_{0}^{L}\rightarrow KK}  &  =\frac{1}{8\pi m_{f_{0}^{L%
}}}\sqrt{1-\left(  \frac{2m_{K}}{m_{f_{0}^{L}}}\right)  ^{2}}\left\vert
\mathcal{M}_{f_{0}^{L}\rightarrow KK}\right\vert ^{2},
\end{align}
where the matrix elements, using the notations $H_{N}\equiv\frac{1}%
{4}\left(  g_{1}^{2}+2h_{1}+h_{2}\right)  $ and $H_{S}\equiv\frac{1}%
{2}\left(  g_{1}^{2}+h_{1}+h_{2}\right)  $, are
\begin{align}
  \mathcal{M}_{f_{0}^{L}\rightarrow KK} & =-\sin\theta_{\sigma
  }\mathcal{M}_{f_{0}K}^{H}(m_{f_{0}^{L}})+\cos\theta_{\sigma
  }\mathcal{M}_{f_{0}K}^{L}(m_{f_{0}^{L}}),\\
  \mathcal{M}_{f_{0}^{H}\rightarrow KK} & =\cos\theta_{\sigma
  }\mathcal{M}_{f_{0}K}^{H}(m_{f_{0}^{H}})+\sin\theta_{\sigma
  }\mathcal{M}_{f_{0}K}^{L}(m_{f_{0}^{H}}),\\
  \mathcal{M}_{f_{0}K}^{L}(m) &
  =-Z_{K}^{2}\bigg[(2\lambda_{1}+\lambda_{2})\phi_{N}
  -\frac{\lambda_{2}}{\sqrt{2}} \phi_{S} + g_{1} w_{K_{1}}
  (m_{K}^{2}-m^{2})\nom \\
  & + w_{K_{1}}^{2}\left( 2H_{N}\phi_{N}-\frac{h_{3}-g_{1}^{2}}
    {\sqrt{2}}\phi_{S}\right)\frac{m^{2}-2m_{K}^{2}}{2} - \frac{c_1}{2}\bigg], \label{cc}\\
  \mathcal{M}_{f_{0}K}^{H}(m) & =-Z_{K}^{2}\bigg[ 2(\lambda_{1}+\lambda_{2})\phi_{S} - 
  \frac{\lambda_{2}}{\sqrt{2}}\phi_{N}+\sqrt{2} g_{1}w_{K_{1}}(m_{K}^{2}-m^{2}) \nom \\
  & \quad +
  w_{K_{1}}^{2}\left(2H_{S}\phi_{S}-\frac{h_{3}-g_{1}^{2}}{\sqrt{2}}\phi_{N}\right)
  \frac{m^{2} - 2m_{K}^{2}}{2}\bigg]. \label{dd}
\end{align}
In the $f_0^{L/H}$ decays only the $\mathcal{M}_{f_{0}\pi}^{L}(m)$ and
$\mathcal{M}_{f_{0}K}^{L}(m)$ expressions have changed.

\section{Experimental data and fitting results for the parametrizations}
\label{App:param_data}

In this appendix we give all the experimental data used for the
determination of the parameters. With the exception of the constituent
quark masses for which we use the values from Chap.~5.5 of
Ref.~\cite{Griffiths:2008zz} (see \secref{Sec:parametrization} as
well), the data are taken from the PDG \cite{PDG} with some necessary
modifications explained in detail in \cite{elsm_2013}. Some of the
data were not used in \cite{elsm_2013} or were used differently there;
these are the following: $m_{a_0}$, $m_{f_{0}(500)}$, $m_{f_{0}(980)}$,
$\Gamma_{a_{0}(980)}$, $\Gamma_{f_{0}(500)\rightarrow\pi\pi}$,
$\Gamma_{f_{0}(500)\rightarrow KK}$,
$\Gamma_{f_{0}(980)\rightarrow\pi\pi}$, and $\Gamma_{f_{0}(980)\rightarrow
  KK}$, for which the values are taken from the PDG. In general we
allowed for larger errors than the ones in the PDG, namely, $20\%$ for
the scalar sector, $10\%$ for the constituent quarks, and $5\%$ for
everything else. However, if for a quantity the PDG error turned out
to be larger, then we used the error value from the PDG.

\begin{table*}[t]
  \caption{
    Experimental values of masses and decay widths and best fit results in
    three cases: original fit of \cite{elsm_2013} that is without fitting
    ${f_0}$ masses, without $\Gamma_{f_0^{L/H}\rightarrow\pi\pi/KK}$ decay
    widths, and with the fermions excluded from the model (left ``Fit''
    column); best and second best solutions with the current approach
    explained in \secref{Sec:parametrization} (middle and right ``Fit''
    columns, respectively). The labels in the first row refer in order to
    the particle assignments of $a_0$, $K_0^{\star},$ and the two $f_0$'s
    (with low and high mass).  We use the \emph{scientific E notation} in
    which $m\e{-n}$ corresponds to $m\times 10^{-n}.$
  }
\label{Tab:observ_particle_setup}
\centering
\begin{tabular} 
[c]{|c||c|c||c|c||c|c|}\hline 
Observable                               &  $\text{Exp.}^{2,2,\text{N},\text{N}}$ [GeV]    &   $\text{Fit}^{2,2,\text{N},\text{N}}$ [GeV]    & $\text{Exp.}^{1,1,1,2}$ [GeV] & $\text{Fit}^{1,1,1,2}$ [GeV]  & $\text{Exp.}^{1,1,1,3}$ [GeV] & $\text{Fit}^{1,1,1,3}$ [GeV] \\\hline             
$f_{\pi}$                                &  $9.221\e{-2} \pm 1.6\e{-4}$  &   $9.630\e{-2}$               & $9.221\e{-2} \pm 1.6\e{-4}$   & $9.55\e{-2}$                  & $9.221\e{-2} \pm 1.6\e{-4}$   & $9.420\e{-2}$                \\\hline  
$f_{K}$      				 &  $0.1105 \pm 8.0\e{-4}$       &   $0.1069$                    & $0.1105 \pm 8.0\e{-4}$        & $0.1094$		         & $0.1105 \pm 8.0\e{-4}$        & $0.1095$                     \\\hline	
$m_{\pi}$			         &  $0.1380  \pm 3.0\e{-3}$      &   $0.1410$                    & $0.1380  \pm 3.0\e{-3}$       & $0.1405$ 			 & $0.1380  \pm 3.0\e{-3}$       & $0.1392$                     \\\hline	
$m_{\eta}$   			         &  $0.54786 \pm 1.8\e{-5}$      &   $0.5094$                    & $0.54786 \pm 1.8\e{-5}$       & $0.5421$			 & $0.54786 \pm 1.8\e{-5}$       & $0.5473$                     \\\hline	
$m_{\eta^{\prime}}$ 			 &  $0.95778 \pm 6.0\e{-5}$      &   $0.9625$                    & $0.95778 \pm 6.0\e{-5}$       & $0.9643$			 & $0.95778 \pm 6.0\e{-5}$       & $0.9595$                     \\\hline	
$m_{K}$      			         &  $0.49564 \pm 2.0\e{-3}$  	 &   $0.4856$                    & $0.49564 \pm 2.0\e{-3}$       & $0.4995$			 & $0.49564 \pm 2.0\e{-3}$       & $0.5076$                     \\\hline	
$m_{\rho}$      		         &  $0.7753 \pm 3.4\e{-4}$ 	 &   $0.7831$                    & $0.7753 \pm 3.4\e{-4}$        & $0.8064$		         & $0.7753 \pm 3.4\e{-4}$        & $0.8021$                     \\\hline	   
$m_{\phi}$    			         &  $1.019461 \pm 1.9\e{-5}$ 	 &   $0.9751$                    & $1.019461 \pm 1.9\e{-5}$      & $0.9901$       		 & $1.019461 \pm 1.9\e{-5}$      & $1.0026$                     \\\hline		
$m_{K^{\star}}$  			 &  $0.8947 \pm 3.0\e{-4}$ 	 &   $0.8851$                    & $0.8947 \pm 3.0\e{-4}$        & $0.9152$			 & $0.8947 \pm 3.0\e{-4}$        & $0.9200$                     \\\hline	        
$m_{a_{1}}$  			         &  $1.2300 \pm 4.0\e{-2}$ 	 &   $1.186$                     & $1.2300 \pm 4.0\e{-2}$        & $1.0766$			 & $1.2300 \pm 4.0\e{-2}$        & $1.0773$                     \\\hline		
$m_{f_{1}(1420)}$ 			 &  $1.4264 \pm 9.0\e{-4}$       &   $1.373$                     & $1.4264 \pm 9.0\e{-4}$        & $1.4160$			 & $1.4264 \pm 9.0\e{-4}$        & $1.4282$                     \\\hline		
$m_{a_{0}}$ 			         &  $1.4740 \pm 1.9\e{-2}$ 	 &   $1.363$                     & $0.9800 \pm 2.0\e{-2}$        & $0.7208$			 & $0.9800 \pm 2.0\e{-2}$        & $0.7656$                     \\\hline		 
$m_{K_{0}^{\star}}$ 			 &  $1.4250 \pm 5.0\e{-2}$  	 &   $1.450$                     & $0.682  \pm  2.9\e{-2}$       & $0.7529$			 & $0.682  \pm  2.9\e{-2}$       & $0.8108$                     \\\hline	
$m_{f_0^L}$      			 &  Not used                     &   No fit                      & $0.475  \pm 7.5\e{-2}$        & $0.2837$	                 & $0.475  \pm 7.5\e{-2}$        & $0.2813$                     \\\hline	
$m_{f_0^H}$    			         &  Not used                     &   No fit                      & $0.990  \pm  2.0\e{-2}$       & $0.7376$ 	                 & $1.350  \pm 0.15$             & $0.8024$                     \\\hline	
$m_{u,d}$ 			         &  Not used                     &   No fit                      & $0.308 \pm 3.1\e{-2}$         & $0.3224$		         & $0.308 \pm 3.1\e{-2}$         & $0.3191$                     \\\hline	
$m_s$        			         &  Not used                     &   No fit                      & $0.483 \pm 4.9\e{-2}$         & $0.4577$		         & $0.483 \pm 4.9\e{-2}$         & $0.4513$                     \\\hline	
$\Gamma_{\rho\rightarrow\pi\pi}$         &  $0.1491 \pm 1.1\e{-3}$       &   $0.1609$                    & $0.1491 \pm 1.1\e{-3}$        & $0.1515$  			 & $0.1491 \pm 1.1\e{-3}$        & $0.1505$                     \\\hline	
$\Gamma_{\phi\rightarrow \bar{K}K}$      &  $3.545\e{-3} \pm 2.6\e{-5}$  &   $3.340\e{-3}$               & $3.545\e{-3} \pm 2.6\e{-5}$   & $3.534\e{-3}$ 		 & $3.545\e{-3} \pm 2.6\e{-5}$   & $3.546\e{-3}$                \\\hline	
$\Gamma_{K^{\star}\rightarrow K\pi}$     &  $4.8\e{-2} \pm 1.3\e{-3}$    &   $4.460\e{-2}$               & $4.8\e{-2} \pm 1.3\e{-3}$     & $4.777\e{-2}$ 		 & $4.8\e{-2} \pm 1.3\e{-3}$     & $4.780\e{-2}$                \\\hline	
$\Gamma_{a_{1}\rightarrow\pi\gamma}$     &  $6.40\e{-4}  \pm 2.46\e{-4}$ &   $6.600\e{-4}$               & $6.40\e{-4}  \pm 2.46\e{-4}$  & $3.670\e{-4}$		 & $6.40\e{-4}  \pm 2.46\e{-4}$  & $3.220\e{-4}$                \\\hline	
$\Gamma_{a_{1}\rightarrow\rho\pi}$       &  $0.425  \pm 0.175$           &   $0.5490$                    & $0.425  \pm 0.175$            & $0.1994$     		 & $0.425  \pm 0.175$            & $0.2919$                     \\\hline	
$\Gamma_{f_{1}^H\rightarrow K^{\star}K}$ &  $4.45\e{-2} \pm 2.1\e{-3}$   &   $4.46\e{-2}$                & $4.45\e{-2} \pm 2.1\e{-3}$    & $4.451\e{-2}$		 & $4.45\e{-2} \pm 2.1\e{-3}$    & $4.451\e{-2}$                \\\hline	
$\Gamma_{a_{0}}$                         &  $0.265 \pm 1.3\e{-2}$        &   $0.2660$                    & $7.5\e{-2} \pm 2.5\e{-2}$     & $6.834\e{-2}$		 & $7.5\e{-2} \pm 2.5\e{-2}$     & $7.488\e{-2}$                \\\hline	
$\Gamma_{K_{0}^{\star}\rightarrow K\pi}$ &  $0.27  \pm 8.0\e{-2}$        &   $0.2850$                    & $0.547 \pm 2.4\e{-2}$         & $0.6001$      		 & $0.547 \pm 2.4\e{-2}$         & $0.5515$                     \\\hline	      
$\Gamma_{f_0^L\rightarrow\pi\pi}$        &  Not used                     &   No fit                      & $0.55  \pm 0.15$              & $0.5542$      		 & $0.55  \pm 0.15$              & $0.5526$                     \\\hline  
$\Gamma_{f_0^L\rightarrow K K}$          &  Not used                     &   No fit                      & $0.0   \pm 0.1$               & $0.0$        		 & $0.0   \pm 0.10$              & $0.0$                        \\\hline  
$\Gamma_{f_0^H\rightarrow\pi\pi}$        &  Not used                     &   No fit                      & $7.0\e{-2} \pm 3.0\e{-2}$     & $8.166\e{-2}$    	         & $0.25  \pm 0.10$              & $0.2495$                     \\\hline	
$\Gamma_{f_0^H\rightarrow K K}$          &  Not used                     &   No fit                      & $0.0   \pm 2.0\e{-2}$         & $0.0$ 		         & $0.150 \pm 0.10$              & $0.0$                        \\\hline	
$T_c (\mu_B=0)$                 	 &  Not used                     &   No fit                      & $0.151 \pm 1.51\e{-2}$        & $0.1704$ 		         & $0.151 \pm 1.51\e{-2}$        & $0.1678$                     \\\hline	
\end{tabular}
\end{table*}

The value of different quantities in the pseudoscalar and
(axial)vector sector can be found in
Table~\ref{Tab:observ_particle_setup}. Since that table contains only
a few from the many possible assignments of the scalar particles to
the states of the scalar nonet, we list below all the values of scalar
masses and decay widths used in the fit,
\begin{eqnarray}
  m_{a_{0}(980)} &=& (980 \pm 20)~\text{MeV},\nom\\
  \Gamma_{a_{0}(980)} &=& (75 \pm 25)~\text{MeV},\nom\\
  m_{a_{0}(1450)} &=& (1474 \pm 19)~\text{MeV},\nom\\
  \Gamma_{a_{0}(1450)} &=& (265 \pm 13)~\text{MeV},\nom\\
  m_{K_{0}^{\star}(800)} &=& (682 \pm 29)~\text{MeV},\nom\\
  \Gamma_{K_{0}^{\star}(800) \rightarrow K\pi} &=& (547 \pm 24)~\text{MeV},\nom\\ 
  m_{K_{0}^{\star}(1430)} &=& (1425 \pm 50)~\text{ MeV},\nom\\
  \Gamma_{K_{0}^{\star}(1430) \rightarrow K\pi} &=& (270 \pm 80)~\text{MeV},\nom\\
  m_{f_{0}(500)} &=& (475 \pm 75)~\text{MeV},\nom\\
  \Gamma_{f_{0}(500)\rightarrow\pi\pi} &=& (550 \pm 150 )~\text{MeV},\nom\\
  \Gamma_{f_{0}(500)\rightarrow KK}  &=& (0 \pm 100 )~\text{MeV},\nom\\
  m_{f_{0}(980)} &=& (990 \pm 20)~\text{MeV},\nom\\  
  \Gamma_{f_{0}(980)\rightarrow\pi\pi}  &=& (70 \pm 30 )~\text{MeV},\nom\\
  \Gamma_{f_{0}(980)\rightarrow KK}  &=& (0 \pm 20 )~\text{MeV},\nom\\
  m_{f_{0}(1370)} &=& (1350 \pm 150)~\text{MeV},\nom\\
  \Gamma_{f_{0}(1370)\rightarrow\pi\pi}  &=& (250 \pm 100)~\text{MeV},\nom\\
  \Gamma_{f_{0}(1370)\rightarrow KK}   &\approx& (150 \pm 100)~\text{MeV},\nom\\ 
  m_{f_{0}(1500)} &=& (1505 \pm 6)~\text{MeV},\nom\\
  \Gamma_{f_{0}(1500)\rightarrow\pi\pi} &=& (38 \pm 2.6)~\text{MeV},\nom\\
  \Gamma_{f_{0}(1500)\rightarrow KK} &=& (9.4\pm1.9)~\text{MeV},\nom\\
  m_{f_{0}(1710)} &=& (1722 \pm 6)~\text{MeV},\nom\\
  \Gamma_{f_{0}(1710)\rightarrow\pi\pi}  &=& (29.3 \pm 5)~\text{MeV},\nom \\
  \Gamma_{f_{0}(1710)\rightarrow KK} &=& (71.4 \pm 18)~\text{MeV}.
\end{eqnarray}


\pagebreak


\begin{thebibliography}{99}

\bibitem{elsm_2013} 
  D.~Parganlija, P.~Kov\'acs, G.~Wolf, F.~Giacosa and D.~H.~Rischke,
  Phys.\ Rev.\ D {\bf 87}, 014011 (2013).

\bibitem {PDG} 
  J.~Beringer {\it et al.}  [Particle Data Group Collaboration],
  Phys.\ Rev.\ D {\bf 86}, 010001 (2012).
  

\bibitem{Chen:2007xr} 
  H.~X.~Chen, A.~Hosaka and S.~L.~Zhu,
  Phys.\ Rev.\ D {\bf 76}, 094025 (2007).


\bibitem{Chen:2009gs} 
  H.~X.~Chen, A.~Hosaka, H.~Toki and S.~L.~Zhu,
  Phys.\ Rev.\ D {\bf 81}, 114034 (2010).


\bibitem{Kojo:2008hk} 
  T.~Kojo and D.~Jido,
  Phys.\ Rev.\ D {\bf 78}, 114005 (2008).


\bibitem{Schaefer:2008hk} 
  B.~J.~Schaefer and M.~Wagner,
  Phys.\ Rev.\ D {\bf 79}, 014018 (2009).


\bibitem{Chatterjee:2011jd} 
  S.~Chatterjee and K.~A.~Mohan,
  Phys.\ Rev.\ D {\bf 85}, 074018 (2012).


\bibitem{Giacosa:2006tf} 
  F.~Giacosa,
  Phys.\ Rev.\ D {\bf 75}, 054007 (2007).


\bibitem{Skokov:2010sf} 
  V.~Skokov, B.~Friman, E.~Nakano, K.~Redlich and B.-J.~Schaefer,
  Phys.\ Rev.\ D {\bf 82}, 034029 (2010).


\bibitem{Gupta:2011ez} 
  U.~S.~Gupta and V.~K.~Tiwari,
  Phys.\ Rev.\ D {\bf 85}, 014010 (2012).


\bibitem{Schaefer:2011ex} 
  B.~J.~Schaefer and M.~Wagner,
  Phys.\ Rev.\ D {\bf 85}, 034027 (2012).


\bibitem{Eser:2015pka}
  J.~Eser, M.~Grahl and D.~H.~Rischke,
  Phys.\ Rev.\ D {\bf 92}, 096008 (2015).


\bibitem{Struber:2007bm} 
  S.~Str\"uber and D.~H.~Rischke,
  Phys.\ Rev.\ D {\bf 77}, 085004 (2008).


\bibitem{Cornwall:1974vz}
  J.~M.~Cornwall, R.~Jackiw and E.~Tomboulis,
  Phys.\ Rev.\ D {\bf 10}, 2428 (1974).


\bibitem{Tawfik:2014gga} 
  A.~N.~Tawfik and A.~M.~Diab, 
  Phys.\ Rev.\ C {\bf 91}, 015204 (2015).


\bibitem{Geffen_1969}
  S.~Gasiorowicz and D.~A.~Geffen,
  Rev.\ Mod.\ Phys.\ \textbf{41}, 531 (1969).


\bibitem{Kaymakcalan_1984}
  O.~Kaymakcalan and J.~Schechter,
  Phys.\ Rev.\ D {\bf 31} (1985) 1109.


\bibitem{Ko_1994}
  P.~Ko and S.~Rudaz,
  Phys.\ Rev.\ D \textbf{50}, 6877 (1994).


\bibitem{U1A_analysis} 
  P.~Kov\'acs and G.~Wolf,
  Acta Phys.\ Polon.\ Supp.\  {\bf 6}, 853 (2013).


\bibitem{Holland:2000uj} 
  K.~Holland and U.~J.~Wiese,
  in {\it At the Frontier of Particle Physics}, edited by M. Shifman
  (World Scientific, Singapore, 2001), Vol. 3, pp. 1909-1944.


\bibitem{Fukushima:2003fw} 
  K.~Fukushima,
  Phys.\ Lett.\ B {\bf 591}, 277 (2004).


\bibitem{Ratti:2005jh} 
  C.~Ratti, M.~A.~Thaler and W.~Weise,
  Phys.\ Rev.\ D {\bf 73}, 014019 (2006).


\bibitem{Hansen:2006ee} 
  H.~Hansen, W.~M.~Alberico, A.~Beraudo, A.~Molinari, M.~Nardi and C.~Ratti,
  Phys.\ Rev.\ D {\bf 75}, 065004 (2007).


\bibitem{Pisarski_2000} 
  R.~D.~Pisarski,
  Phys.\ Rev.\ D {\bf 62}, 111501 (2000).


\bibitem{Scavenius:2002ru} 
  O.~Scavenius, A.~Dumitru and J.~T.~Lenaghan,
  Phys.\ Rev.\ C {\bf 66}, 034903 (2002).


\bibitem{Sasaki_2006} 
  C.~Sasaki, B.~Friman and K.~Redlich,
  Phys.\ Rev.\ D {\bf 75}, 074013 (2007).


\bibitem{Roessner_2006} 
  S.~Roessner, C.~Ratti and W.~Weise,
  Phys.\ Rev.\ D {\bf 75}, 034007 (2007).


\bibitem{Haas:2013qwp} 
  L.~M.~Haas, R.~Stiele, J.~Braun, J.~M.~Pawlowski and J.~Schaffner-Bielich,
  Phys.\ Rev.\ D {\bf 87}, 076004 (2013).


\bibitem{Borsanyi:2012ve}
  S.~Bors\'anyi, G.~Endr\H{o}di, Z.~Fodor, S.~D.~Katz and K.~K.~Szab{\'o},
  JHEP {\bf 1207}, 056 (2012).


\bibitem{Schaefer:2007pw}
  B.~J.~Schaefer, J.~M.~Pawlowski and J.~Wambach,
  Phys.\ Rev.\ D {\bf 76}, 074023 (2007).


\bibitem{Reinosa:2015gxn} 
  U.~Reinosa, J.~Serreau, M.~Tissier and N.~Wschebor,
  Phys.\ Rev.\ D {\bf 93}, 105002 (2016).


\bibitem{Kapusta:2006pm} J.~I.~Kapusta and C.~Gale, {\it
    Finite-Temperature Field Theory: Principles and Applications}
  (Cambridge University Press, Cambridge, UK, 2006).


\bibitem{Tiwari:2013pg} 
  V.~K.~Tiwari,
  Phys.\ Rev.\ D {\bf 88}, 074017 (2013).


\bibitem{Gupta:2009fg} 
  U.~S.~Gupta and V.~K.~Tiwari,
  Phys.\ Rev.\ D {\bf 81}, 054019 (2010).


\bibitem{Mintz:2012mz}
  B.~W.~Mintz, R.~Stiele, R.~O.~Ramos and J.~Schaffner-Bielich,
  Phys.\ Rev.\ D {\bf 87}, 036004 (2013).


\bibitem{Rossner:2007ik}
  S.~Roessner, T.~Hell, C.~Ratti and W.~Weise,
  Nucl.\ Phys.\ A {\bf 814}, 118 (2008).


\bibitem{Reinosa:2015oua}
  U.~Reinosa, J.~Serreau and M.~Tissier,
  Phys.\ Rev.\ D {\bf 92}, 025021 (2015).


\bibitem{MINUIT} 
  F.~James and M.~Roos,
  Comput.\ Phys.\ Commun.\  {\bf 10} (1975) 343.


\bibitem{giacosa_2010} 
  F.~Giacosa and G.~Pagliara,
  Nucl.\ Phys.\ A {\bf 833}, 138 (2010).


\bibitem{Griffiths:2008zz}
  D.~Griffiths,
  {\it Introduction to elementary particles},
  Weinheim, Germany: Wiley-VCH (2008).


\bibitem{Aoki:2006br} 
  Y.~Aoki, Z.~Fodor, S.~D.~Katz and K.~K.~Szabo,
  Phys.\ Lett.\ B {\bf 643}, 46 (2006).


\bibitem{Borsanyi:2010bp}
  S.~Bors\'anyi, Z.~Fodor, C.~Hoelbling, S.~D.~Katz, S.~Krieg, C.~Ratti,
  K.~K.~Szab{\'o},
  JHEP {\bf 1009}, 073 (2010).
  

\bibitem{Cheng:2007jq} 
  M.~Cheng {\it et al.},
  Phys.\ Rev.\ D {\bf 77}, 014511 (2008).


\bibitem{Megias:2004hj} 
  E.~Megias, E.~Ruiz Arriola and L.~L.~Salcedo,
  Phys.\ Rev.\ D {\bf 74}, 065005 (2006).


\bibitem{Oleszczuk:1992yg} 
  M.~Oleszczuk and J.~Polonyi,
  Annals Phys.\  {\bf 227}, 76 (1993).


\bibitem{Costa:2005cz} 
  P.~Costa, M.~C.~Ruivo, C.~A.~de Sousa and Y.~L.~Kalinovsky,
  Phys.\ Rev.\ D {\bf 71}, 116002 (2005).


\bibitem{Kunihiro:1989my} 
  T.~Kunihiro,
  Phys.\ Lett.\ B {\bf 219}, 363 (1989).


\bibitem{Ruivo:2012xt} 
  M.~C.~Ruivo, P.~Costa and C.~A.~de Sousa,
  Phys.\ Rev.\ D {\bf 86}, 116007 (2012).


\bibitem{Marko:2013lxa} 
  G.~Mark{\'o}, U.~Reinosa and Z.~Sz{\'e}p,
  Phys.\ Rev.\ D {\bf 87}, 105001 (2013).


\bibitem{Andersen:2015sma}
  J.~O.~Andersen, T.~Brauner and W.~Naylor,
  Phys.\ Rev.\ D {\bf 92}, 114504 (2015).
  
  
\bibitem{Borsanyi:2010cj}
  S.~Bors\'anyi, G.~Endr\H{o}di, Z.~Fodor, A.~Jakov\'ac, S.~D.~Katz,
  S.~Krieg, C.~Ratti and K.~K.~Szab{\'o},
  JHEP {\bf 1011}, 077 (2010).


\bibitem{Skokov:2010wb}
  V.~Skokov, B.~Stokic, B.~Friman and K.~Redlich,
  Phys.\ Rev.\ C {\bf 82}, 015206 (2010).
 

\bibitem{Marko_2010} 
  G.~Mark{\'o} and Zs.~Sz{\'e}p,
  Phys.\ Rev.\ D {\bf 82}, 065021 (2010).


\bibitem{Cea:2015cya}
  P.~Cea, L.~Cosmai and A.~Papa,
  Phys.\ Rev.\ D {\bf 93}, 014507 (2016).


\bibitem{Bonati:2015bha}
  C.~Bonati, M.~D'Elia, M.~Mariti, M.~Mesiti, F.~Negro and F.~Sanfilippo,
  Phys.\ Rev.\ D {\bf 92}, 054503 (2015).


\bibitem{Bellwied:2015rza}
  R.~Bellwied, S.~Bors\'anyi, Z.~Fodor, J.~G\"unther, S.~D.~Katz, C.~Ratti and
  K.~K.~Szab{\'o},
  Phys.\ Lett.\ B {\bf 751}, 559 (2015).


\bibitem{BraunMunzinger:2003zz}
  P.~Braun-Munzinger, J.~Stachel and C.~Wetterich,
  Phys.\ Lett.\ B {\bf 596}, 61 (2004).
  

\bibitem{Cleymans:2006qe}
  J.~Cleymans, H.~Oeschler, K.~Redlich and S.~Wheaton,
  J.\ Phys.\ G {\bf 32}, S165 (2006).


\bibitem{Stiele:2016cfs} 
  R.~Stiele and J.~Schaffner-Bielich,
  Phys.\ Rev.\ D {\bf 93}, 094014 (2016).


\bibitem{Mao:2009aq} 
  H.~Mao, J.~Jin and M.~Huang,
  J.\ Phys.\ G {\bf 37}, 035001 (2010).

  
\bibitem{Gavai:2004sd}
  R.~V.~Gavai and S.~Gupta,
  Phys.\ Rev.\ D {\bf 71}, 114014 (2005).


\bibitem{deForcrand:2006pv}
  P.~de Forcrand and O.~Philipsen,
  JHEP {\bf 0701}, 077 (2007).


\bibitem{Chen:2009gv}
  J.~W.~Chen, K.~Fukushima, H.~Kohyama, K.~Ohnishi and U.~Raha,
  Phys.\ Rev.\ D {\bf 80}, 054012 (2009).


\bibitem{Jakovac:2010uy}
  A.~Jakov\'ac and Zs.~Sz{\'e}p,
  Phys.\ Rev.\ D {\bf 82}, 125038 (2010).


\bibitem{Qin:2010nq}
  S.~x.~Qin, L.~Chang, H.~Chen, Y.~x.~Liu and C.~D.~Roberts,
  Phys.\ Rev.\ Lett.\ {\bf 106}, 172301 (2011).


\bibitem{Shi:2014zpa}
  C.~Shi, Y.~L.~Wang, Y.~Jiang, Z.~F.~Cui and H.~S.~Zong,
  JHEP {\bf 1407}, 014 (2014).


\bibitem{Gutierrez:2013sta}
  E.~Gutierrez, A.~Ahmad, A.~Ayala, A.~Bashir and A.~Raya,
  J.\ Phys.\ G {\bf 41}, 075002 (2014).


  
\bibitem{Fischer:2014ata} 
  C.~S.~Fischer, J.~Luecker and C.~A.~Welzbacher,
  Phys.\ Rev.\ D {\bf 90}, 034022 (2014).


\bibitem{Eichmann:2015kfa} 
  G.~Eichmann, C.~S.~Fischer and C.~A.~Welzbacher,
  Phys.\ Rev.\ D {\bf 93}, 034013 (2016).


\bibitem{Ayala:2011vs}
  A.~Ayala, A.~Bashir, C.~A.~Dominguez, E.~Gutierrez, M.~Loewe and A.~Raya,
  Phys.\ Rev.\ D {\bf 84}, 056004 (2011).


\bibitem{Lacey:2014wqa}
  R.~A.~Lacey,
  Phys.\ Rev.\ Lett.\  {\bf 114}, 142301 (2015).

\bibitem{Pelaez:2015qba} 
  J.~R.~Pelaez,
  arXiv:1510.00653 [hep-ph].

\bibitem{Wolkanowski:2015lsa} 
  T.~Wolkanowski, F.~Giacosa and D.~H.~Rischke,
  Phys.\ Rev.\ D {\bf 93}, 014002 (2016).

\bibitem{Wolkanowski:2015jtc} 
  T.~Wolkanowski, M.~Soltysiak and F.~Giacosa,
  Nucl. Phys. {\bf B909}, 418 (2016).

\bibitem{Fariborz:2015era} 
  A.~H.~Fariborz, A.~Azizi and A.~Asrar,
  Phys.\ Rev.\ D {\bf 91}, 073013 (2015).

\bibitem{Janowski:2011gt} 
  S.~Janowski, D.~Parganlija, F.~Giacosa and D.~H.~Rischke,
  Phys.\ Rev.\ D {\bf 84}, 054007 (2011).

\bibitem{Janowski:2014ppa} 
  S.~Janowski, F.~Giacosa and D.~H.~Rischke,
  Phys.\ Rev.\ D {\bf 90}, 114005 (2014).

\end{thebibliography}
\end{document}